\documentclass[twocolumn,10pt,5p]{elsarticle}
\usepackage[hyperindex,breaklinks]{hyperref} 
\usepackage{preprintcover}  
\usepackage{atlasphysics}
\usepackage{subfigure}
\usepackage{multirow}
\usepackage{amsmath}
\usepackage{amssymb}
\usepackage{lineno}
\usepackage{ifpdf}
\ifpdf
\usepackage{grffile}

\else

\fi
\usepackage{subfigure}
\usepackage{multirow}
\usepackage{amsmath}
\usepackage{amssymb}
\usepackage{hypernat}
\usepackage{lineno}

\newcommand\progname[1]{{\sc #1}}

\newcommand\ztoll{$Z\to\ell\ell$}
\newcommand\htolvlv{$H\to W^+W^-\to\ell^+\nu\ell^-\bar{\nu}$}
\newcommand\htollll{$H\to ZZ\to\ell^+\ell^-\ell^+\ell^-$}
\newcommand\htollvv{$H\to ZZ\to\ell^+\ell^-\nu\bar{\nu}$}
\newcommand\htottvv{$H\to ZZ\to\tau^+\tau^-\nu\bar{\nu}$}

\newcommand\zztollvv{$ZZ\to\ell^+\ell^-\nu\bar{\nu}$}
\newcommand\htollqq{$H\to ZZ\to\ell^+\ell^- q \bar{q}$}

\newcommand\totallumi{4.7~${\rm fb}^{-1}$}
\newcommand\totallumisummer{1.0~${\rm fb}^{-1}$}
\newcommand\totallumiupd{2.0~${\rm fb}^{-1}$}

\def\met{\ensuremath{E_{\mathrm{T}}^{\mathrm{miss}}}}

\def\mH{\ensuremath{m_{\mathrm{H}}}}
\def\pt{\ensuremath{p_{\mathrm{T}}}}

\def\progname#1{{\sc #1}}

\newif\ifnotpagetest
\notpagetesttrue

\ifnotpagetest
\else
\fi

\def\papertitle{
Search for a Standard Model Higgs boson in the \htollvv\ decay channel using \totallumi~of $\sqrt{s}=7$~TeV data with the ATLAS Detector
}

\def\theabstract{%
A search for a heavy Standard Model Higgs~boson decaying via \htollvv, where $\ell$ represents electrons or muons, is presented.  
It is based on proton-proton collision data at $\sqrt{s}=7$~TeV, collected by the ATLAS experiment at the LHC
during 2011 and corresponding to an integrated luminosity of \totallumi. 
The data agree with the expected Standard Model backgrounds.  
Upper limits on the Higgs~boson production cross section are derived for Higgs boson masses between 200 GeV and 600 GeV  
and the production of a Standard Model Higgs boson with a mass in the range $319-558\:$~GeV is excluded at the 95\% confidence level. 
}

\PreprintCoverPaperTitle{\papertitle}  % paper title
\PreprintIdNumber{CERN-PH-EP-2012-120}  % CERN preprint number
\PreprintCoverAbstract{\theabstract}  % paper abstract
\PreprintJournalName{Physics Letters B}  % journal name
\journal{Physics Letters B}

\begin{document}
%\linenumbers

\begin{frontmatter}
\title{\papertitle}
\author{The ATLAS Collaboration}
\address{}
\begin{abstract}
\ifnotpagetest
\theabstract
\fi
\end{abstract}
\end{frontmatter}

\section{Introduction}

The search for the Higgs boson~\cite{Englert:1964et,Higgs:1964pj,Guralnik:1964eu}, which in the Standard Model (SM) gives mass to the 
weak vector bosons and other particles, 
is one of the most important aspects of the CERN Large Hadron Collider (LHC) physics programme. 
Direct searches performed at the CERN Large Electron-Positron Collider (LEP) have excluded, at a 95\% confidence level (CL), 
the production of a SM Higgs boson with mass, \mH, less than $114.4\:$~GeV~\cite{Barate:2003sz}. 
Searches at the Fermilab Tevatron $p\bar{p}$ collider have excluded at 95\% CL the region $156<\mH < 177\:$~GeV~\cite{Tevatron,tevhiggs}. 
At the LHC, the combination of ATLAS searches~\cite{Collaboration:2012si}, using $1.0-4.9$~fb$^{-1}$ of $\sqrt{s}=7\:$~TeV 
data, 
excluded at the 95\% CL the production of a SM Higgs boson in the regions  $112.9<\mH <115.5\:$~GeV, $131<\mH <238\:$~GeV and  $251<\mH <466\:$~GeV. 
The CMS combined result~\cite{Chatrchyan:2012tx} using $4.6-4.8$~fb$^{-1}$ of $\sqrt{s}=7\:$~TeV data
excluded the production of a SM Higgs boson at the 95\% CL in the region $127<\mH < 600\:$~GeV.

The \zztollvv\ decay channel offers a substantial branching fraction in combination with a good separation from potential 
background processes owing to the large transverse momentum, $p_{\mathrm{T}}$, of the electron or muon pair from the leptonic $Z$ boson decay and 
the large missing transverse momentum from the $Z$~boson decaying to neutrinos. 
Earlier results in this channel, published in Ref.~\cite{Aad:2011va}, using \totallumisummer~were subsequently updated in Ref.~\cite{Collaboration:2012si} with \totallumiupd~and  
exclude at the 95\% CL the presence of a SM Higgs boson in the range $310<\mH <470\:$~GeV. The most recent CMS analysis~\cite{Chatrchyan:2012vv} in this channel based on $4.6$~fb$^{-1}$ excluded $270<\mH <440\:$~GeV.

The data sample considered in the search presented in this Letter was recorded by the ATLAS experiment during the 2011 LHC run at a centre-of-mass energy $\sqrt{s}=7\:$~TeV.
The integrated luminosity of the data sample, considering only data-taking periods where all relevant detector subsystems were operational, is \totallumi~with an 
uncertainty of 3.9\%~\cite{Aad:2011dr,lumi2011}.

\section{The ATLAS detector}
The ATLAS detector~\cite{Aad:2008zzm} is a multi-purpose particle physics detector with forward-backward symmetric cylindrical 
geometry\footnote{ATLAS uses a right-handed coordinate system with its origin at the nominal interaction point. 
The $z$-axis is along the beam pipe, the $x$-axis points to the centre of the LHC ring and the $y$-axis points upward. 
Cylindrical coordinates ($r$,$\phi$) are used in the transverse plane, $\phi$ being the azimuthal angle around the beam pipe. 
The pseudorapidity $\eta$ is defined as $\eta=-\ln[\tan(\theta/2)]$ where $\theta$ is the polar angle.}. 
The inner tracking detector (ID) 
covers $\left|\eta\right|<2.5$ and consists of a silicon pixel detector, 
a silicon microstrip detector, and a transition radiation tracker. 
The ID is surrounded by a thin superconducting solenoid providing a $2\:\rm{T}$ axial magnetic field. 
A high-granularity lead/liquid-argon (LAr) sampling calorimeter 
measures the energy and the position of electromagnetic showers with $\left|\eta\right|<3.2$. 
LAr sampling calorimeters are also used to measure hadronic showers in the end-cap ($1.5<\left|\eta\right|<3.2$) and forward ($3.1<\left|\eta\right|<4.9$) regions, 
while an iron/scintillating-tile calorimeter 
measures hadronic showers in the central region ($\left|\eta\right|<1.7$). 
The muon spectrometer 
surrounds the calorimeters and consists of three large superconducting air-core toroids, 
each with eight coils, a system of precision tracking chambers ($\left|\eta\right|<2.7$), and fast tracking chambers for triggering ($\left|\eta\right|<2.4$). 
A three-level trigger system 
selects events to be recorded for offline analysis. 

\section{Data and simulation samples}
The \htollvv\ signal is modelled using the \progname{powheg}~\cite{powheg4,powheg5} event generator, which includes matrix elements for the gluon fusion and vector-boson 
fusion (VBF) production mechanisms of the Higgs~boson up to next-to-leading-order (NLO). \progname{powheg} is interfaced to \progname{pythia}~\cite{pythia} for the modelling of parton showers. 
The modelling of final-state radiation is performed with \progname{photos}~\cite{Golonka:2005pn}, and \progname{tauola}~\cite{Jadach:1993hs,Golonka:2003xt} is used for the simulation of $\tau$ decays. 
The Higgs boson \pt~spectrum in the gluon fusion process is reweighted to the calculation of Ref.~\cite{deFlorian:2011xf}, which provides QCD corrections up to NLO and QCD soft-gluon resummations 
up to next-to-next-to-leading logarithms (NNLL). 
 
\htolvlv, \htollqq\ and \htollll\ samples are also simulated using the same generators as for the \htollvv\ samples. 
These channels, together with the \htottvv\ final state, which is included in the \htollvv\ samples, contribute 
to the signal yield and are considered as part of the signal.
In particular, \htolvlv\ decays contribute as much as 70\% to the signal expectation after the full selection for $\mH=200$~GeV, decreasing to 13\% at $\mH=300$~GeV, and to less than 4.5\% for Higgs boson masses larger than 400~GeV.
Orthogonality of the analysis with respect to other ATLAS Higgs boson searches~\cite{ATLAS:2011aa,Aad:2011ec,Collaboration:2012sm} 
is ensured through mutually exclusive selection requirements on the dilepton invariant mass, the number of leptons and the missing transverse momentum in the event.

The Higgs boson production cross sections and decay branching ratios, 
as well as their uncertainties, are taken from Refs.~\cite{LHCHiggsCrossSectionWorkingGroup:2011ti,LHCHiggsCrossSectionWorkingGroup:2012vm}. 
The cross sections for the gluon fusion process have been calculated to NLO in QCD in Refs.~\cite{Djouadi:1991tka,Dawson:1990zj,Spira:1995rr}, 
and to next-to-next-to-leading-order (NNLO) in Refs.~\cite{Harlander:2002wh,Anastasiou:2002yz,Ravindran:2003um}. 
In addition, QCD soft-gluon resummations, calculated in the NNLL approximation~\cite{Catani:2003zt}, are applied for the gluon fusion process. 
NLO electroweak (EW) corrections are also applied~\cite{Aglietti:2004nj,Actis:2008ug}. 
These results are compiled in Refs.~\cite{Anastasiou:2008tj,deFlorian:2009hc,Baglio:2010ae} and assume factorisation between QCD and EW corrections. 
The cross sections for VBF processes are calculated with full NLO QCD and EW corrections~\cite{Ciccolini:2007jr,Ciccolini:2007ec,Arnold:2008rz}, 
and approximate NNLO QCD corrections are included~\cite{Bolzoni:2010xr}.
The uncertainty in the production cross section due to the choice of QCD scale is typically $^{+12}_{-8}\%$ for the gluon fusion process, and
$\pm 1\%$ for the VBF process~\cite{LHCHiggsCrossSectionWorkingGroup:2011ti}. 
The uncertainty in the production cross section due to the parton distribution function (PDF) and $\alpha_{s}$ is $\pm 8\%$ for gluon-initiated processes and $\pm 4\%$ 
for quark-initiated processes~\cite{Botje:2011sn,Lai:2010vv,Martin:2009iq,Ball:2011mu}. 
The Higgs boson decay branching ratio~\cite{Djouadi:1997yw} to the four-fermion final state is predicted by \progname{prophecy4f}~\cite{Bredenstein:2006rh,Bredenstein:2006ha}.

The Higgs boson cross sections are calculated with a zero-width approximation. For the Higgs decay, an ad~hoc 
relativistic Breit-Wigner line shape is applied at the event-generator level. 
It has been suggested~\cite{Seymour:1995qg,Passarino:2010qk,Anastasiou:2011pi,LHCHiggsCrossSectionWorkingGroup:2012vm} that effects related to off-shell Higgs~boson production 
and interference with other SM processes may become sizeable for the highest Higgs boson masses ($\mH>400\:$~GeV) considered in this search. 
Currently, in the absence of a full lineshape calculation for the production mechanisms as well as a correct account of  
the interference with SM $ZZ$ production, a conservative estimate of the possible size of such effects is included as a signal normalisation 
systematic uncertainty, following a parameterisation as a function of $\mH$: $1.5\times[\mH/\mathrm{TeV}]^3$, for $\mH\ge300\:$~GeV~\cite{LHCHiggsCrossSectionWorkingGroup:2012vm}. 

Different event generators are chosen to model a range of important background processes. 
The \progname{alpgen} generator~\cite{Mangano:2002ea} interfaced to \progname{herwig}~\cite{Corcella:2000bw} for parton showers and
hadronisation is used to simulate inclusive $W/Z$~boson backgrounds. 
\progname{mc@nlo}~\cite{Frixione:2002ik}, interfaced to \progname{herwig} and \progname{jimmy}~\cite{jimmy}, 
is used for the production of top-pair, single-top and diboson ($WW$, $WZ$ and $ZZ$) backgrounds. 
\progname{pythia} is used to simulate $b\bar{b}$ and $c\bar{c}$ samples as well as alternative samples for the $Z$ and $ZZ$ backgrounds. 
All simulated background samples are scaled to the highest-precision calculations available for the relevant process. 
An overview of the relevant predictions and their uncertainties is given in Ref.~\cite{Aad:2011qv}.

Generated events are simulated using the ATLAS detector simulation~\cite{simuAtlas} within the \progname{geant4} framework~\cite{GEANT4}. 
Additional $pp$ interactions in the same and nearby bunch crossings (pile-up) are included in the simulation. 
The Monte Carlo (MC) samples are reweighted to reproduce the observed distribution of the number of interactions per bunch crossing in the data.

Data used for the search in the muon channel are collected using a single muon trigger
with a $p_{\mathrm{T}}$ threshold of 18~GeV, while in the electron case a logical OR between a single electron trigger with a threshold varying from 20 to 22~GeV and 
a dielectron trigger with a $p_{\mathrm{T}}$ threshold of 12~GeV is used.
The expected trigger efficiency is close to 100\% in the electron channel and between 94\% and 97\% in the muon channel for 
signal events passing the full selection cuts described below.

During the course of the 2011 data-taking the average number of pile-up interactions increased due to increased beam currents and stronger beam focusing. 
This changed the average number of interactions per bunch crossing at the start of a fill, from about six in the earlier periods to about 15 in the later periods.
The missing transverse momentum resolution is affected by the level of pile-up, resulting in a significant change in the signal-to-background ratio between 
the earlier and the later periods. 
To retain the best sensitivity, the search 
is therefore split  between the earlier (2.3~fb$^{-1}$) and the later (2.4~fb$^{-1}$) periods, hereafter referred to as the ``low pile-up data'' 
and the ``high pile-up data'', respectively. The selection is unaltered between the periods.

\section{Lepton identification and event selection}

Electron candidates consist of clusters of energy deposited in the electromagnetic calorimeter that are associated to ID tracks. 
The electron candidates must satisfy a set of identification criteria~\cite{Aad:2011mk} that require the shower profiles to be 
consistent with those expected for electromagnetic showers and a well-reconstructed ID track pointing to the corresponding cluster. 
Furthermore the electron candidates are required to have $\pt>20\:$~GeV and pseudorapidity $|\eta|<2.47$. 
The electron transverse momentum is computed from the cluster energy and the track direction at the interaction point. 

Muons are identified by reconstructing tracks in the muon spectrometer. 
These tracks are then extrapolated back to the beam line to find a matching ID track. 
Details of muon reconstruction and identification can be found in Ref.~\cite{Aad:2010yt}.
Only muons with $\pt>20\:$~GeV and $|\eta| < 2.5$ are considered.

Jets are used in this analysis to reject backgrounds from events with heavy-quark decays or from events with fake missing transverse momentum \met\ due to mis-measured jets. 
For this purpose, jets are reconstructed from clusters of energy deposits in the calorimeters using the 
anti-$k_{\mathrm{t}}$ algorithm~\cite{AntiKt} with a radius parameter $R=0.4$.
Jets are calibrated using $p_{\rm T}$- and $\eta$-dependent correction factors based on MC simulation and validated with data~\cite{jetcalibration}; 
this calibration corrects for effects of energy
from additional proton-proton interactions.
Only jets with $p_{\mathrm{T}}>25$~GeV and $|\eta|<2.5$ are considered. 

Two conditions are applied to remove leptons associated with jets, such as those originating from semi-leptonic decays of 
$b$-hadrons. Leptons are not considered in the analysis if the scalar sum of ID track momenta, not 
associated with the lepton, in a cone 
$\Delta R < 0.2$ around the lepton direction is greater than $10\%$ of the $p_{\mathrm{T}}$ of the lepton itself or 
if the lepton is within a distance $\Delta R=0.4$ of the nearest jet.

\begin{figure*}
  \centering
  \subfigure[\label{fig:llvv1a}]{\includegraphics[width=0.49\textwidth]{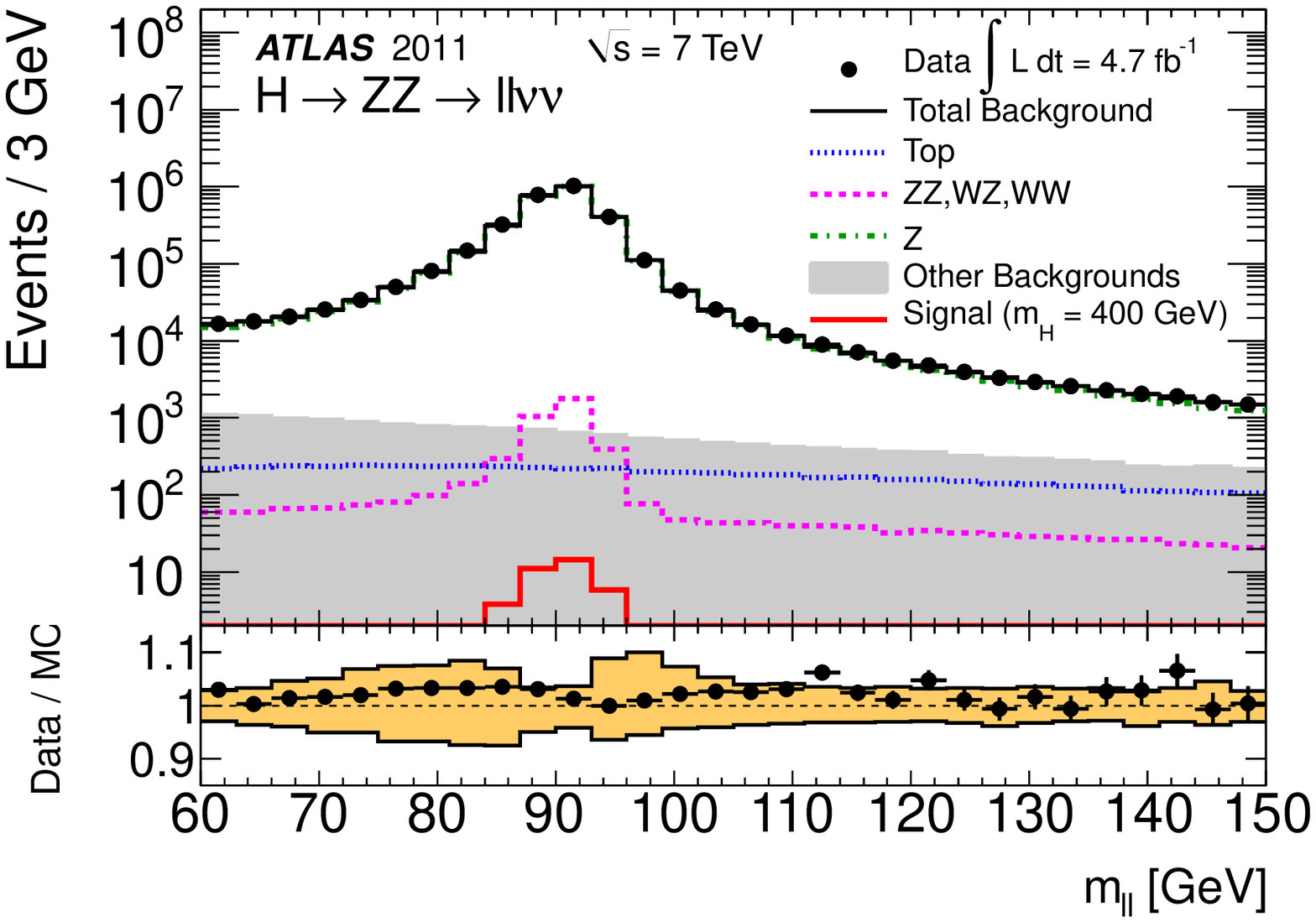}}
  \subfigure[\label{fig:llvv1b}]{\includegraphics[width=0.49\textwidth]{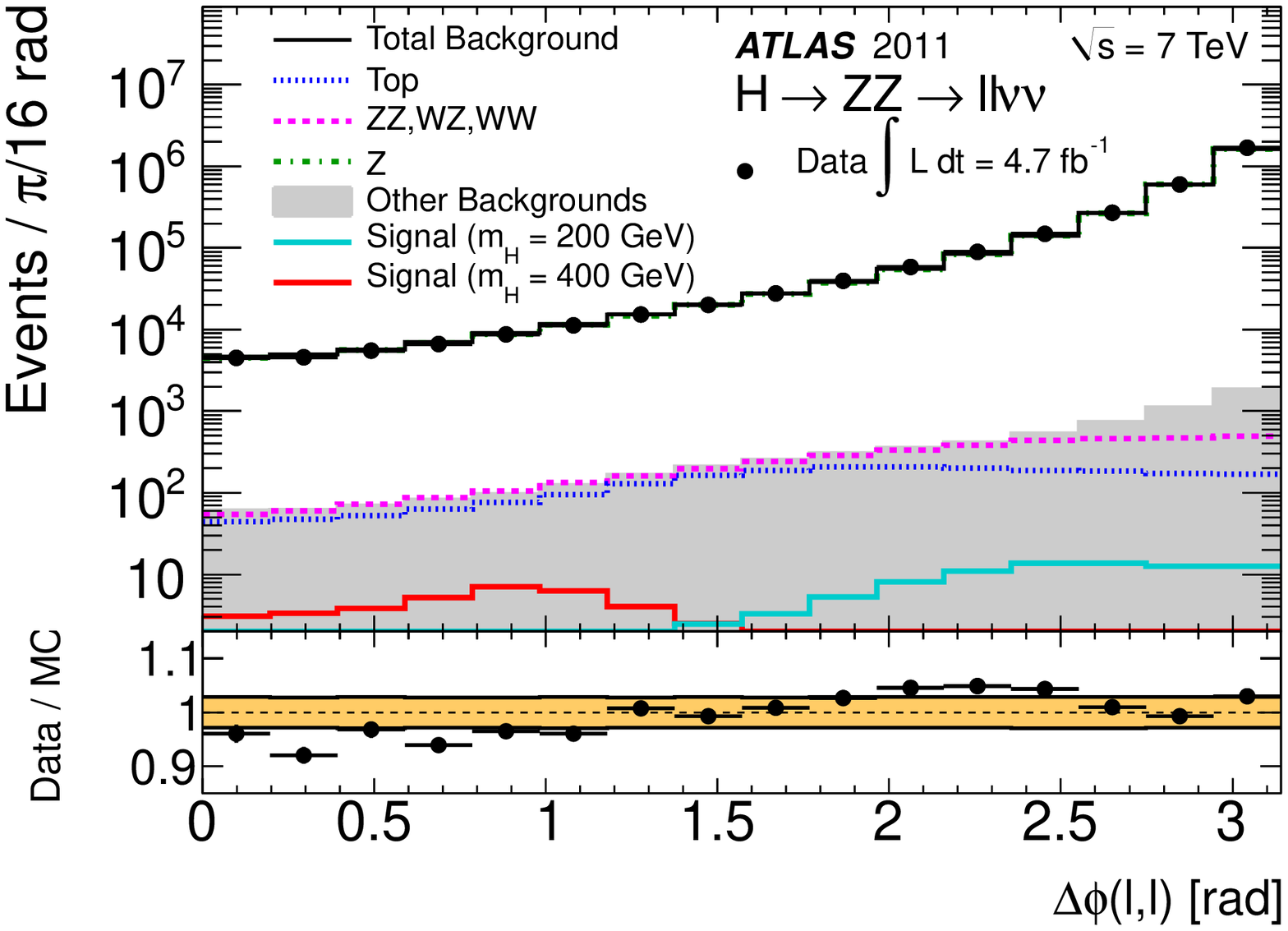}}
  \caption{\subref{fig:llvv1a} The dilepton invariant mass distribution for events with exactly two oppositely-charged electrons or muons. 
    \subref{fig:llvv1b} The azimuthal separation between leptons for events with exactly two oppositely-charged electrons or muons with an invariant mass satisfying 
    $|m_{\mathrm{Z}}-m_{\ell\ell}|<15\:$~GeV.\ The inset at the bottom of the figures shows the ratio of the data to the combined background expectations 
    as well as a band corresponding to the combined systematic uncertainties. In both figures the line corresponding to the $Z$~boson background is mostly hidden 
under the line for the total background. The background contribution labeled ``Other Backgrounds'' includes multijet and inclusive $W$~boson production.  
    \label{fig:htollvvcontrol}}
\end{figure*}

The missing transverse momentum is measured as the negative vectorial sum of the transverse momenta of all 
cells in the calorimeters with $|\eta|<4.9$, calibrated appropriately based on their identification as electrons, photons, $\tau$-leptons, jets or unassociated calorimeter cells, and all selected muons in the event~\cite{met}. 
Calorimeter deposits associated with muons are subtracted to avoid double counting.

\begin{figure*}
  \centering
  \subfigure[\label{fig:2a}]{\includegraphics[width=0.49\textwidth]{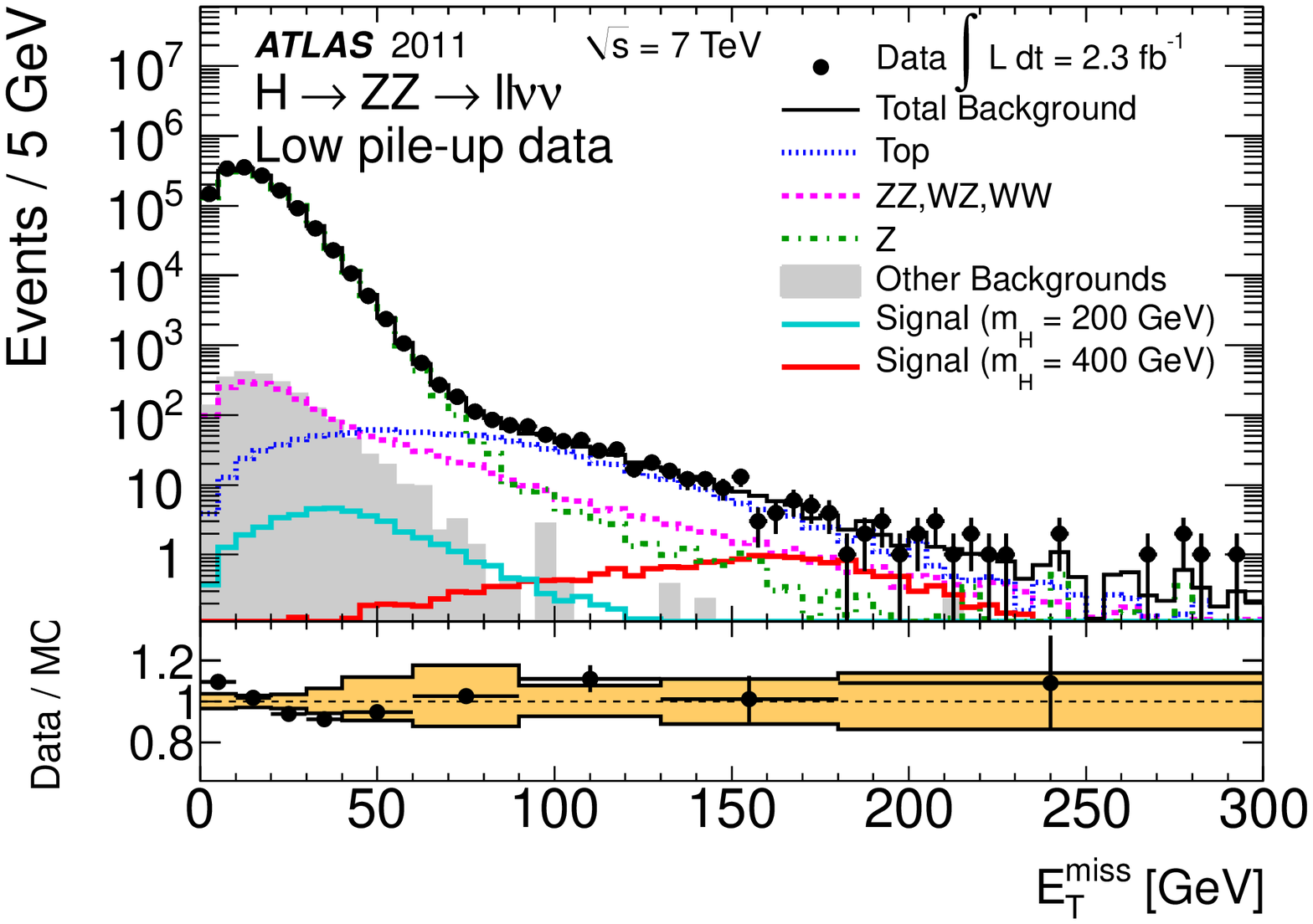}}
  \subfigure[\label{fig:2b}]{\includegraphics[width=0.49\textwidth]{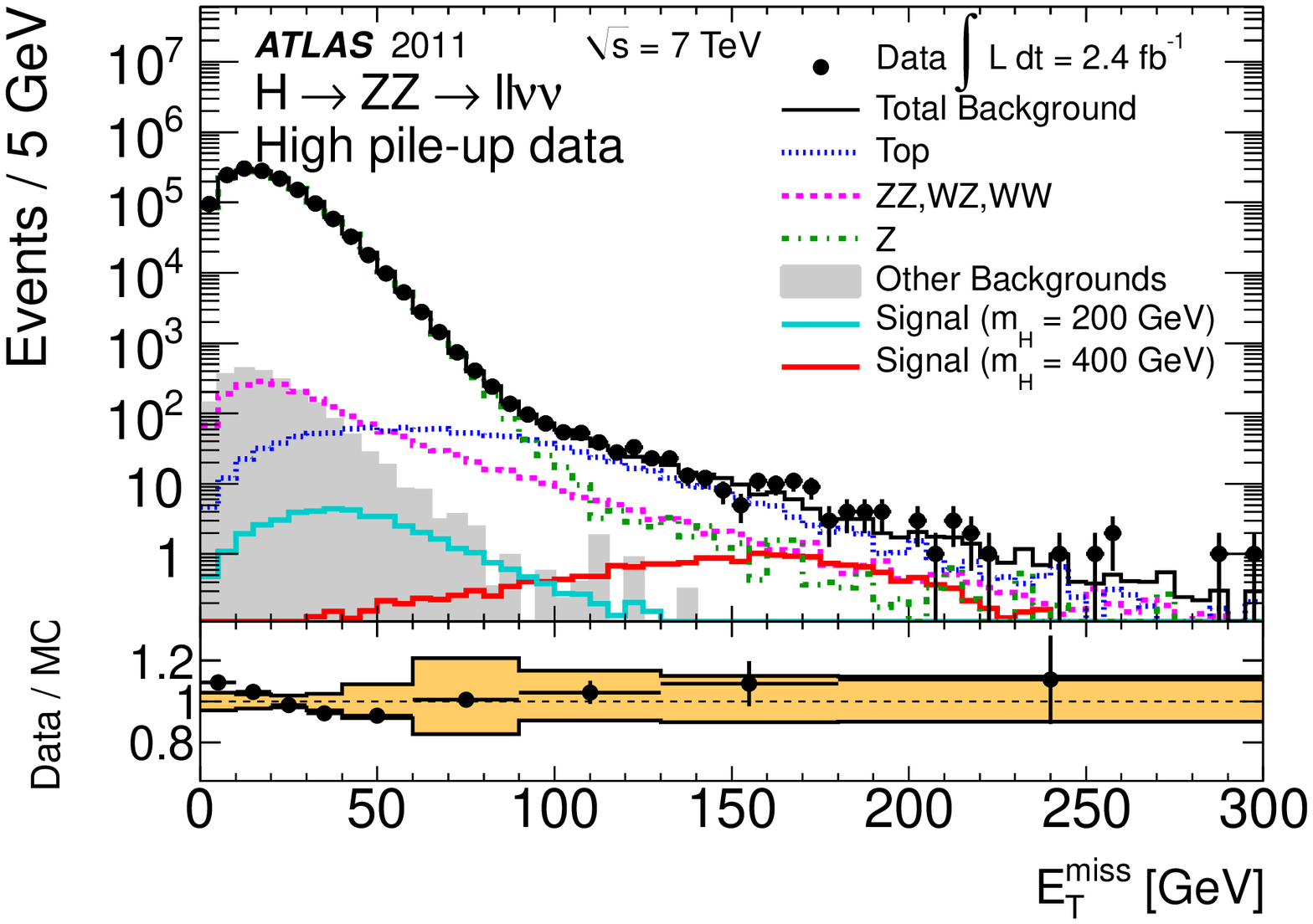}}
  \caption{The \met\ distributions for events with exactly two oppositely-charged electrons or muons satisfying $|m_{\mathrm{Z}}-m_{\ell\ell}|<15\:$~GeV, 
    for  
    \subref{fig:2a} the low pile-up data and \subref{fig:2b} the high pile-up data. 
    The insets at the bottom show the ratio of the data to the combined background expectations as well as a band corresponding to the combined systematic uncertainties. The background contribution labeled ``Other Backgrounds'' includes multijet and inclusive $W$~boson production.
    \label{fig:htollvvcontrol2}}
\end{figure*}

Events are required to contain a reconstructed primary vertex, 
with at least three associated tracks with $p_{\rm T}>0.4$~GeV, and exactly two oppositely-charged electrons or muons, 
consistent with having originated from the primary vertex.  
Furthermore, events are rejected if they contain a third lepton, as defined above, but with a loosened $p_{\mathrm{T}}$ requirement of at least $10\:$~GeV.
The dilepton mass distribution is shown in Fig.~\ref{fig:llvv1a}. 
Inclusive $Z$~boson production is the dominant background at this stage of the analysis. 
To suppress backgrounds from top, inclusive $W$~boson, and multijet production, the dilepton invariant mass, $m_{\ell\ell}$, is required to satisfy $|m_{\mathrm{Z}}-m_{\ell\ell}|<15\:$~GeV.

To reduce the background from top~quark production, events with one or more $b$-tagged jets are rejected, where the $b$-tagging is based on a multivariate algorithm which uses 
information from both the impact parameter with respect to the primary vertex of tracks associated to the jet and the presence of displaced secondary vertices associated to the 
jet's tracks. 
Jets with a loosened $p_{\mathrm{T}}$ threshold of $20\;$~GeV and $\left|\eta\right|<2.5$ are considered in the $b$-tag veto. 
The selection applied on the $b$-tagging discriminant achieves an efficiency of about 85\%(50\%) for identifying $b$-jets($c$-jets)~\cite{btagnew} 
and a light-jet rejection factor of about ten~\cite{mistagnew}, in inclusive top-pair events. 

To exploit the mass-dependent kinematic features of \htollvv\ production, the search is subdivided into a low Higgs~boson mass region
($\mH <280$~GeV) and a high Higgs~boson mass region ($\mH \geq280$~GeV), where dedicated cuts are applied 
to two important discriminating variables used to reduce the background contributions: \met\ and the azimuthal angle between the two leptons, $\Delta\phi(\ell,\ell)$.
Events can contribute to one or both search regions depending on whether they satisfy the cuts applied on these variables. 
Figures~\ref{fig:llvv1b} and~\ref{fig:htollvvcontrol2} show the distributions of these variables after the application of the $m_{\ell\ell}$ window cut.
Since inclusive $Z$~boson production gives rise to a steeply falling \met\ distribution, 
systematic uncertainties on the \met\ reconstruction are particularly important in order to estimate this background correctly. 

\begin{table*}
\centering
\caption{
The expected number of background and signal events along with the observed numbers of candidates in the data, separated into the low and high \mH\ search regions and the low and high pile-up periods. 
The quoted uncertainties are statistical and systematic, respectively. 
\label{tab:htollvv_yield}}
{\small
\begin{tabular}{crrrr}\hline\hline
       &  \multicolumn{2}{c}{low $m_{\mathrm{H}}$ search}       & \multicolumn{2}{c}{high $m_{\mathrm{H}}$ search}  \\  \hline
 Source      &  Low pile-up data          &   High pile-up  data    &  Low pile-up data        &   High pile-up  data \\  \hline
  $Z$        &  $40.1\pm 5.0 \pm 7.9$     &  $265\pm 13 \pm 67$ & $0.8\pm 0.3 \pm 0.8$       & $11.6\pm 2.1 \pm 2.9$\\
  $W$        &  $4.6\pm 2.2 \pm 4.6$      &  $5.8\pm 1.8 \pm 5.8$     & $1.5\pm 0.8 \pm 1.5$       & $2.2\pm 1.3 \pm 2.2$\\
  top        &  $23.2\pm 1.3 \pm 5.4$     &  $27.9\pm 1.3 \pm 5.3$    & $16.0\pm 1.1 \pm 4.0$      & $17.2\pm 1.0 \pm 3.9$\\
  multijet   &  $1.1\pm 0.2 \pm 0.5$      &  $1.1\pm 0.2 \pm 0.6$     & $0.1\pm 0.1 \pm 0.0$       & $0.1\pm 0.1 \pm 0.0$\\
  $ZZ$       &  $33.4\pm 0.7 \pm 3.9$     &  $36.7\pm 0.7 \pm 4.3$    & $28.4\pm 0.6 \pm 3.4$      & $31.9\pm 0.7 \pm 3.8$\\
  $WZ$       &  $23.3\pm 1.0 \pm 2.8$     &  $25.2\pm 1.0 \pm 3.0$    & $17.1\pm 0.8 \pm 2.1$      & $18.9\pm 0.8 \pm 2.3$\\
  $WW$       &  $25.5\pm 0.8 \pm 3.0$\    &  $32.4\pm 0.9 \pm 3.8$    & $9.4\pm 0.5 \pm 1.1$       & $13.3\pm 0.5 \pm 1.6$\\
  Total      &  $151 \pm 6  \pm 11$ &  $394\pm 13 \pm 67$ & $73.3 \pm 1.8  \pm 6.1$    & $95.2 \pm 2.9  \pm 6.9$\\
  Data       &         158                &        442                &          77                &    109             \ \\ \hline
    $m_{\mathrm{H}}$ [GeV]  & \multicolumn{4}{c}{Signal expectation }  \\
 \hline
  200        &  $10.3\pm 0.2 \pm 1.8$     &  $11.1\pm 0.2 \pm 1.9$    & &\\
 300         &                            &                           & $16.4\pm 0.3 \pm 2.9$      & $17.5\pm 0.3 \pm 3.1$\\
 400         &                            &                           & $14.4\pm 0.2 \pm 2.5$      & $15.4\pm 0.2 \pm 2.7$\\
 500         &                            &                           & $6.2\pm 0.1 \pm 1.1$       & $6.5\pm 0.1 \pm 1.1$\\
 600         &                            &                           & $2.7\pm 0.0 \pm 0.5$       & $2.9\pm 0.0 \pm 0.5$\\
   \hline\hline
\end{tabular}}
\end{table*}

The dominant contribution to the \met\ uncertainty comes from the knowledge of the jet energy scale. 
A degradation of the \met~resolution is observed in the data taken during the periods 
with a larger average number of interactions per bunch crossing, as is illustrated in Fig.~\ref{fig:htollvvcontrol2}. 
In particular, this increases the background from inclusive $Z$~boson production in the high pile-up periods, mostly in the low mass search region.

Figure~\ref{fig:htollvvcontrol2} shows that the data and the combined background expectation agree within systematic uncertainties.
In the low \mH\ region, events are required to satisfy $\met>66\:$~GeV, whilst in the high \mH\ region the requirement is $\met>82\:$~GeV.
These cuts reduce significantly the backgrounds from processes with no or modest genuine missing transverse momentum originating from unobserved neutrinos. 
In the low \mH\ region a significant fraction of signal events is also removed by this criterion.

The boost of the $Z$~bosons originating from a Higgs~boson decay increases with \mH, thus reducing the expected opening angle between the leptons. 
In the low (high) \mH\ region an upper bound, $\Delta\phi(\ell,\ell) < 2.64~(2.25)$, is therefore applied.
In the low  \mH\ region a lower bound, $\Delta\phi(\ell,\ell) > 1$, is also applied, to reduce backgrounds from events in which 
the lepton pair did not originate from a $Z$ boson decay.
In the high \mH\ region, events are also rejected if the azimuthal angle between the missing transverse momentum vector  
and the direction of the \ztoll~boson candidate is $\Delta\phi(\vec{p}_\mathrm{T}^{\rm\ miss},\vec{p}_\mathrm{T}^{\ \ell\ell}) < 1$.

Finally, to further reduce the background from events with fake \met\ due to mis-measured jets, events are rejected if the azimuthal angle between 
$\vec{p}_\mathrm{T}^{\rm\ miss}$ and the nearest jet in the event satisfies $\Delta\phi(\vec{p}_\mathrm{T}^{\rm\ miss},\vec{p}_\mathrm{T}^{\ \rm jet}) < 1.5$ 
for the low \mH\ search and $\Delta\phi(\vec{p}_\mathrm{T}^{\rm\ miss},\vec{p}_\mathrm{T}^{\ \rm jet}) < 0.5$ for the high \mH\ region.

The efficiency of the event selection is very similar in the electron and muon channels, ranging from 3.1\% for $\mH =200$~GeV to about 43\% for $\mH =600$~GeV.

\section{Background normalisation and control regions}

\begin{figure*}
  \centering
  \subfigure[\label{fig:llvv3a}]{\includegraphics[width=0.49\textwidth]{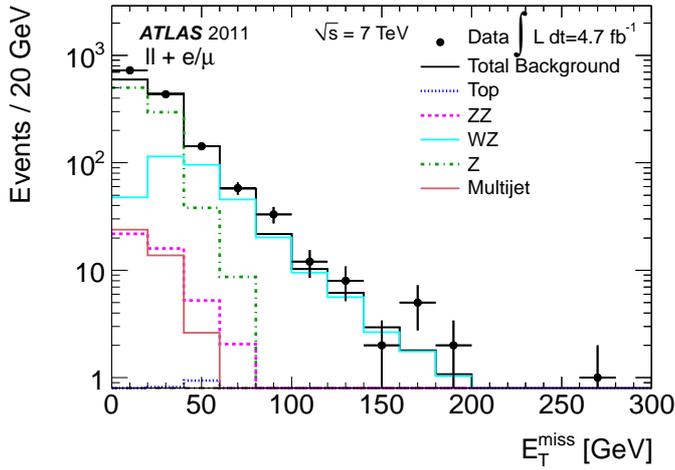}}
  \subfigure[\label{fig:llvv3b}]{\includegraphics[width=0.49\textwidth]{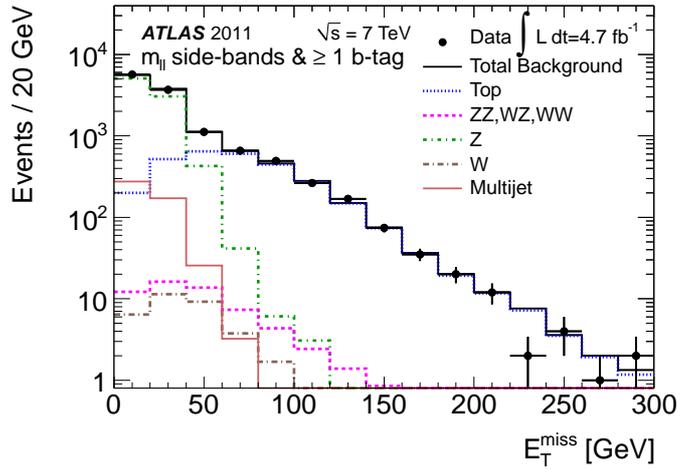}}
  \subfigure[\label{fig:llvv3c}]{\includegraphics[width=0.49\textwidth]{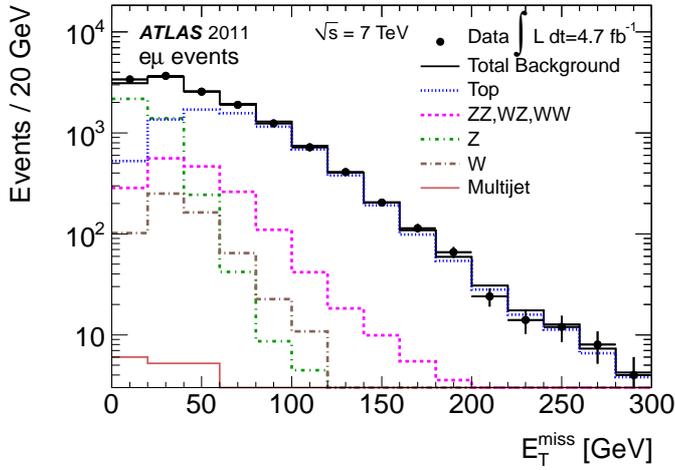}}
  \subfigure[\label{fig:llvv3d}]{\includegraphics[width=0.49\textwidth]{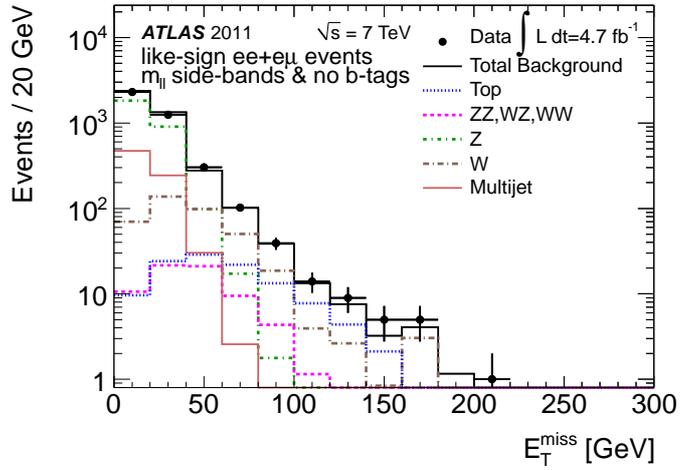}}
  \caption{The \met\ distribution in the different control regions:
    Figure~\subref{fig:llvv3a} is a $WZ$ control region with events containing exactly three leptons.
    Figure~\subref{fig:llvv3b} is a top-quark control region populated with events containing at least one $b$-jet and a lepton pair in the sidebands of dilepton mass distribution.
    Figure~\subref{fig:llvv3c} is a second top-quark control region populated with events with an oppositely-charged $e\mu$ pair.
    Figure~\subref{fig:llvv3d} is an inclusive $W$~boson control region populated with events with no $b$-tagged jets and a same-charge $ee$ or $e\mu$ pair in the 
    sidebands of dilepton mass distribution.
    \label{fig:htollvvcontrol3}}
\end{figure*}

Standard Model pair-production of $Z$~bosons has a final state identical to the signal, and is therefore expected to survive most of the
applied selection criteria and form a continuum in the transverse mass distribution (defined in Section~\ref{sec:results}). 
The normalisation for this background is obtained from a calculation including next-to-leading-order terms~\cite{Campbell:2011bn} 
with an additional 6\% term to account for missing quark-box diagrams ($gg\to  ZZ$)~\cite{Binoth:2008pr}.
An 11\% normalisation uncertainty is assigned to this background, estimated from scale, PDF~\cite{LHCHiggsCrossSectionWorkingGroup:2012vm} and model uncertainties. 
The $ZZ$ background is taken from the MC simulation with a systematic uncertainty given by the shape prediction 
when \progname{pythia} is used as an alternative event generator.
$WW$ and $WZ$ backgrounds are normalised in a similar way. For the $WZ$ background the normalisation is verified using a control sample 
in which the presence of exactly three leptons is required, where the minimum $p_{\mathrm{T}}$ for the third lepton is 10~GeV.
Figure~\ref{fig:llvv3a} shows the \met\ distribution in this control region, which is dominated by $WZ$ background for $\met > 40$~GeV and is well modelled by the 
MC simulations.

The background from top-quark events is taken from the MC prediction. This prediction is verified to agree with data 
in two independent control samples: the first uses the dilepton mass sideband and requires at least one identified $b$-jet (Fig.~\ref{fig:llvv3b}),
while the second selects events containing oppositely-charged electron-muon pairs (Fig.~\ref{fig:llvv3c}).

Additional backgrounds can arise from multijet events or inclusive $W$~boson production due to heavy flavour decays or jets misidentified as leptons. 
The normalisation of the inclusive $W$~boson background in this search is obtained from a control sample of events with a like-sign electron-electron or electron-muon 
pair in the sidebands of the dilepton mass distribution and with no $b$-tagged jets. Figure~\ref{fig:llvv3d} shows the \met\ distribution 
following this procedure. At large \met\ this distribution is dominated by the inclusive $W$~boson background.

The multijet background in the electron channel is determined using a data sample based on a 
loosened electron selection. This sample, which is dominated by jet events, is scaled to describe the tails of the $m_{\ell\ell}$ distribution. 
In the muon channel, the background from heavy flavour decays is studied using simulation, whereas other muon sources
from multijet events are constrained using a sample of like-sign muon pairs in data. 
In both cases the background is found to be negligible.

\begin{figure*}
  \centering
  \subfigure[\label{fig:llvv4b}]{\includegraphics[width=0.49\textwidth]{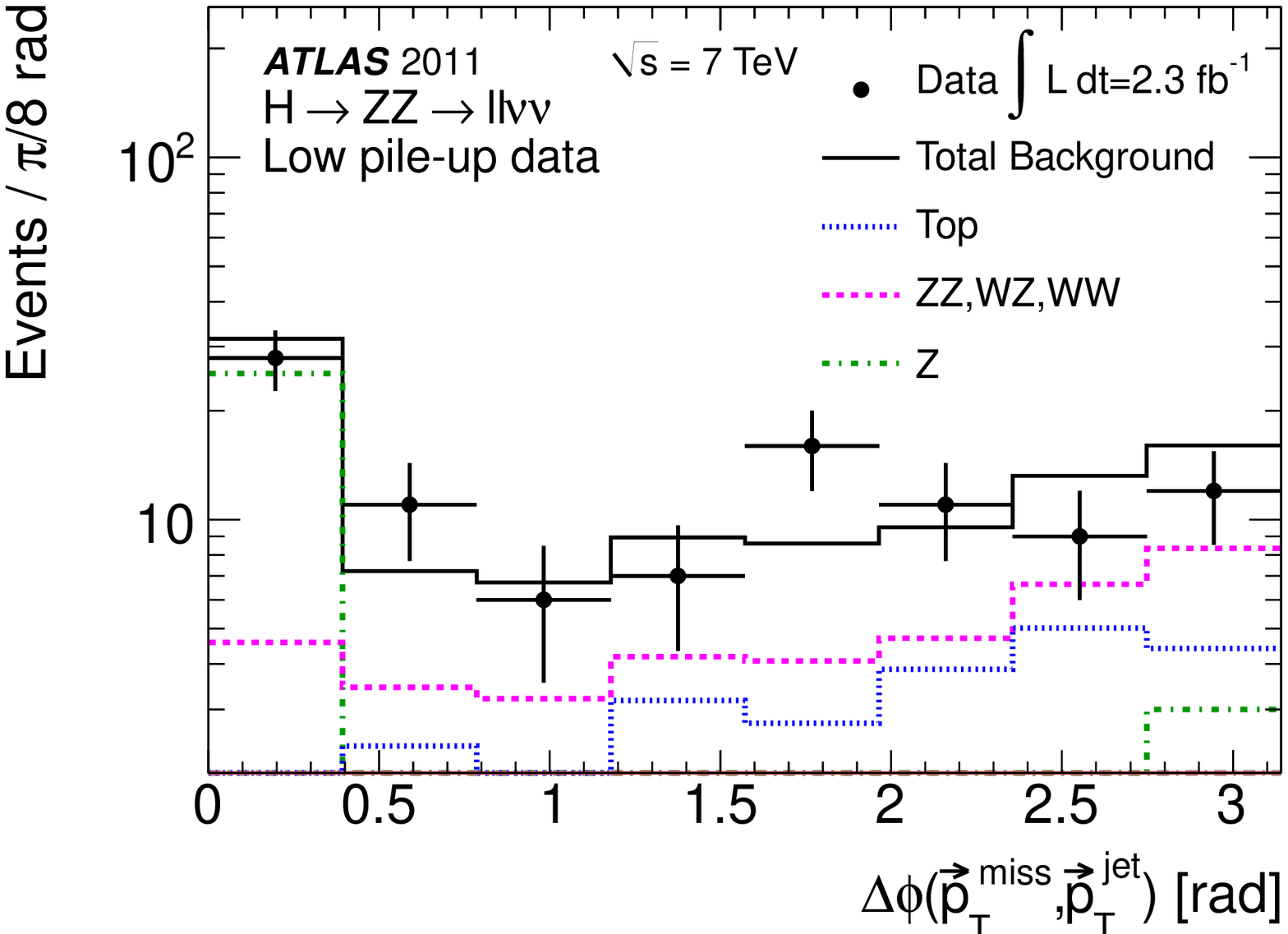}}
  \subfigure[\label{fig:llvv4c}]{\includegraphics[width=0.49\textwidth]{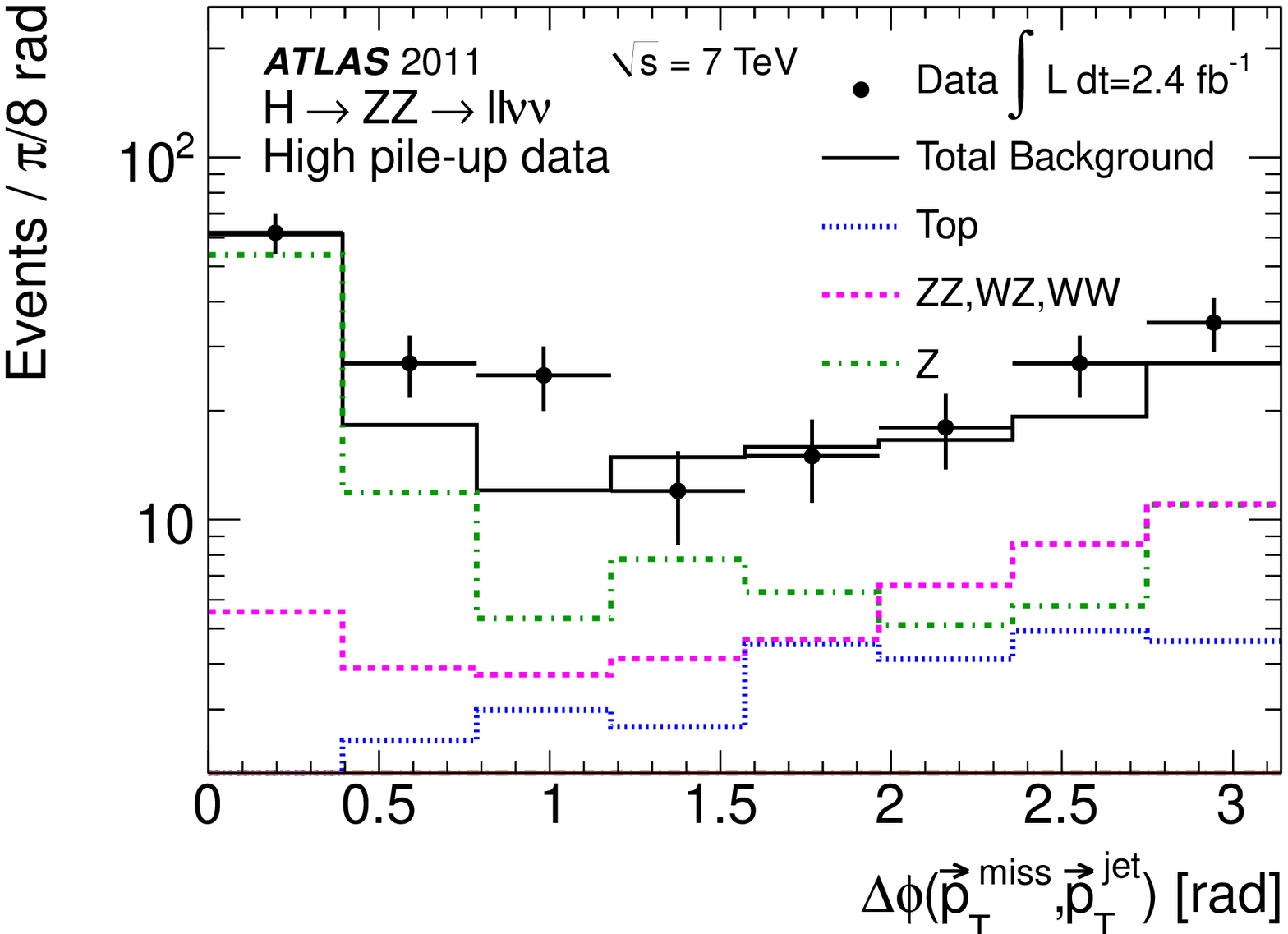}}
  \caption{The azimuthal separation between the missing transverse momentum vector, $\vec{p}_\mathrm{T}^{\rm\ miss}$, and the nearest jet in the event $\Delta\phi(\vec{p}_\mathrm{T}^{\rm\ miss},\vec{p}_\mathrm{T}^{\ \rm jet})$ 
    after the \met\ requirement, for the high \mH\ search region. Figure~\subref{fig:llvv4b} refers to the low pile-up data and figure~\subref{fig:llvv4c} to the high pile-up data.
    \label{fig:htollvvcontrol4}}
\end{figure*}

The background from inclusive $Z$~boson production is taken from the MC prediction and verified in a control region populated with events rejected by the 
$\Delta\phi(\vec{p}_\mathrm{T}^{\rm\ miss},\vec{p}_\mathrm{T}^{\ \rm jet})$ cut  
after the \met\ requirement. The full $\Delta\phi(\vec{p}_\mathrm{T}^{\rm\ miss},\vec{p}_\mathrm{T}^{\ \rm jet})$ distribution is shown in  Fig.~\ref{fig:htollvvcontrol4}. At low $\Delta\phi(\vec{p}_\mathrm{T}^{\rm\ miss},\vec{p}_\mathrm{T}^{\ \rm jet})$, where the inclusive $Z$~boson background dominates, agreement is observed between the data and the expected backgrounds. 

The signal efficiencies and overall background expectations are similar in the electron and muon channels; therefore only combined results are presented.  
The numbers of candidate \htollvv\ events selected in data and the expected yields from signal and background processes are 
shown in Table~\ref{tab:htollvv_yield}. 

\section{Systematic uncertainties}

The systematic uncertainties applied in this search include experimental uncertainties related to the selection and 
calibration of electrons, muons and jets, which are also all explicitly propagated to the \met\ calculation.
Uncertainties on the $b$($c$)-tagging efficiency as well as the light-jet mis-tagging rate are also applied. 

Normalisation uncertainties for the signal (gluon fusion {\scriptsize$^{+14\%}_{-10\%}$} and VBF 4\%)~\cite{LHCHiggsCrossSectionWorkingGroup:2011ti} and 
for the diboson backgrounds (11\%) are obtained from theory; 
uncertainties for the inclusive $Z$~boson production (2.5\%), top~quark production (9\%), inclusive $W$~boson production (100\%) and 
for multijet production in the electron channel (50\%) are estimated from data. Where a normalisation uncertainty is obtained from data, 
additional detector related systematic uncertainties are only considered if they are likely to affect the background in the control region 
and in the signal region differently.

For the signal, an additional uncertainty is applied to account for the possible effect of theoretical uncertainties on the acceptance. 
This uncertainty is estimated from signal samples containing variations in the modelling of initial- and final-state radiation, 
the renormalisation and factorisation scales, which are varied to half and twice their nominal values, as well as variations in the underlying event tune, which was changed from the default \progname{auet2b}~\cite{auet2b} 
to \progname{perugia2011}~\cite{Skands:2010ak}, both based on LHC data.
An overall uncertainty of 8.4\% (3.4\%) is assigned in the low (high) mass search regions, obtained by the addition in quadrature of the 
largest deviations observed for each variation. 

The luminosity uncertainty is 3.7\% and 4.1\% for the low and high pile-up data, respectively, based on the calibration described in Refs.~\cite{Aad:2011dr,lumi2011}. 
Where appropriate, systematic uncertainties are treated as correlated between the signal and the different 
background expectations. 
The total systematic uncertainty on the signal and on each of the background contributions can be seen in Table~\ref{tab:htollvv_yield}.
In most cases the assigned normalisation uncertainties dominate the total systematic uncertainty, except for the inclusive $Z$~boson background, for which  
the jet energy scale and resolution uncertainties dominate, and the top-quark background for which the $b$-tagging uncertainty dominates. In the low \mH\ search using the high pile-up data, the uncertainty on the 
inclusive $Z$~boson background uncertainty dominates the overall uncertainty.

\section{Results}\label{sec:results}

After the event selection, the Higgs~boson search is performed by 
looking for an excess of data over the SM background expectation in the transverse mass distribution of the selected $ee\nu\nu$ and 
$\mu\mu\nu\nu$ events. 
The transverse mass is calculated from the lepton pair and the $\vec{p}_\mathrm{T}^{\rm\ miss}$ vector as:

{\scriptsize \[ m_{\mathrm{T}}^2 \equiv \left[\sqrt{m_{\mathrm{Z}}^2+|\vec{p}_\mathrm{T}^{\ \ell\ell}|^2}+\sqrt{m_{\mathrm{Z}}^2+|\vec{p}_\mathrm{T}^{\rm\ miss}|^2}\right]^2-\left[\vec{p}_\mathrm{T}^{\ \ell\ell}+\vec{p}_\mathrm{T}^{\rm\ miss}\right]^2.\]}

\begin{figure*}
  \centering
  \subfigure[\label{fig:htollvv_mass1}]{\includegraphics[width=0.49\textwidth]{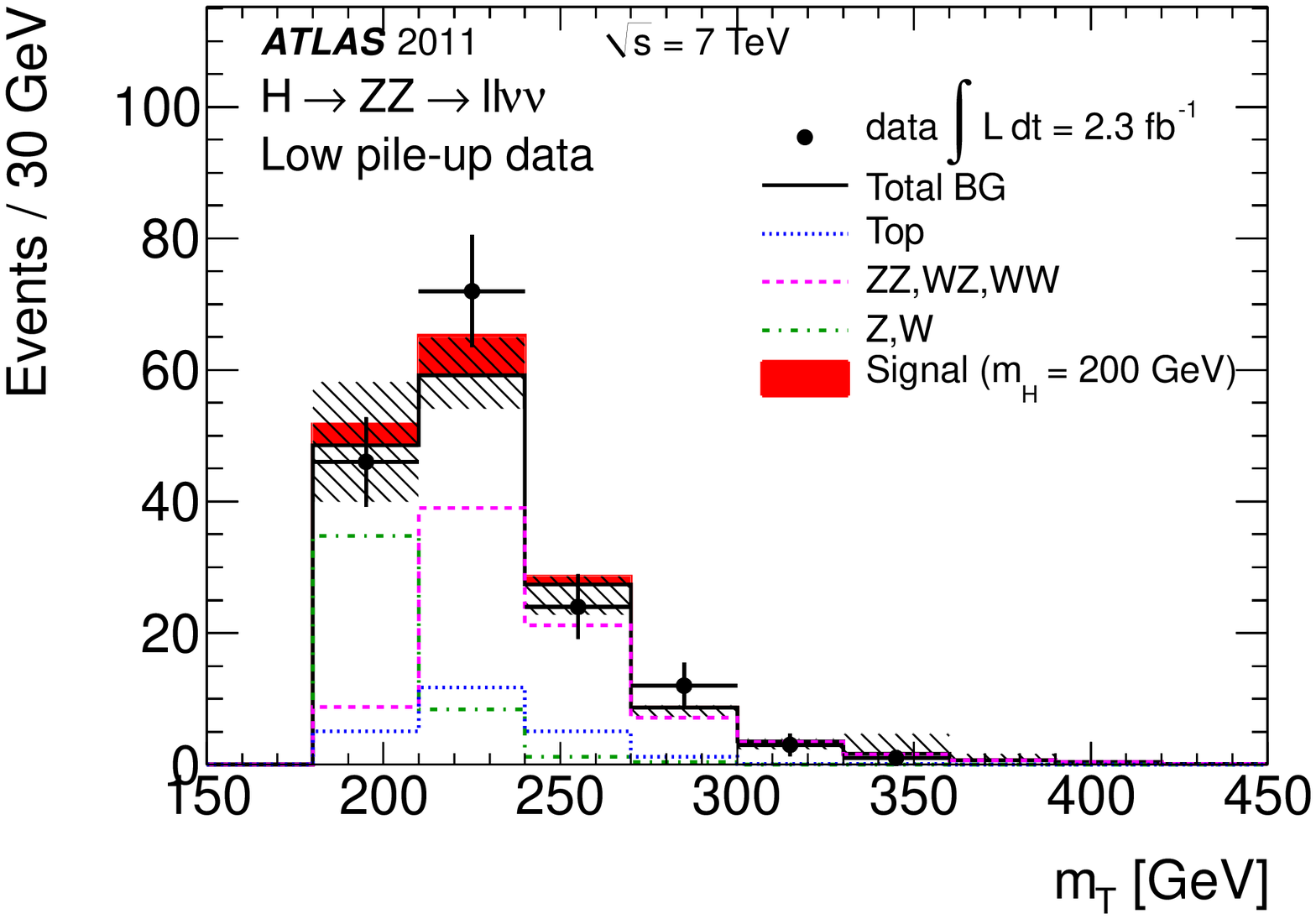}}
  \subfigure[\label{fig:htollvv_mass2}]{\includegraphics[width=0.49\textwidth]{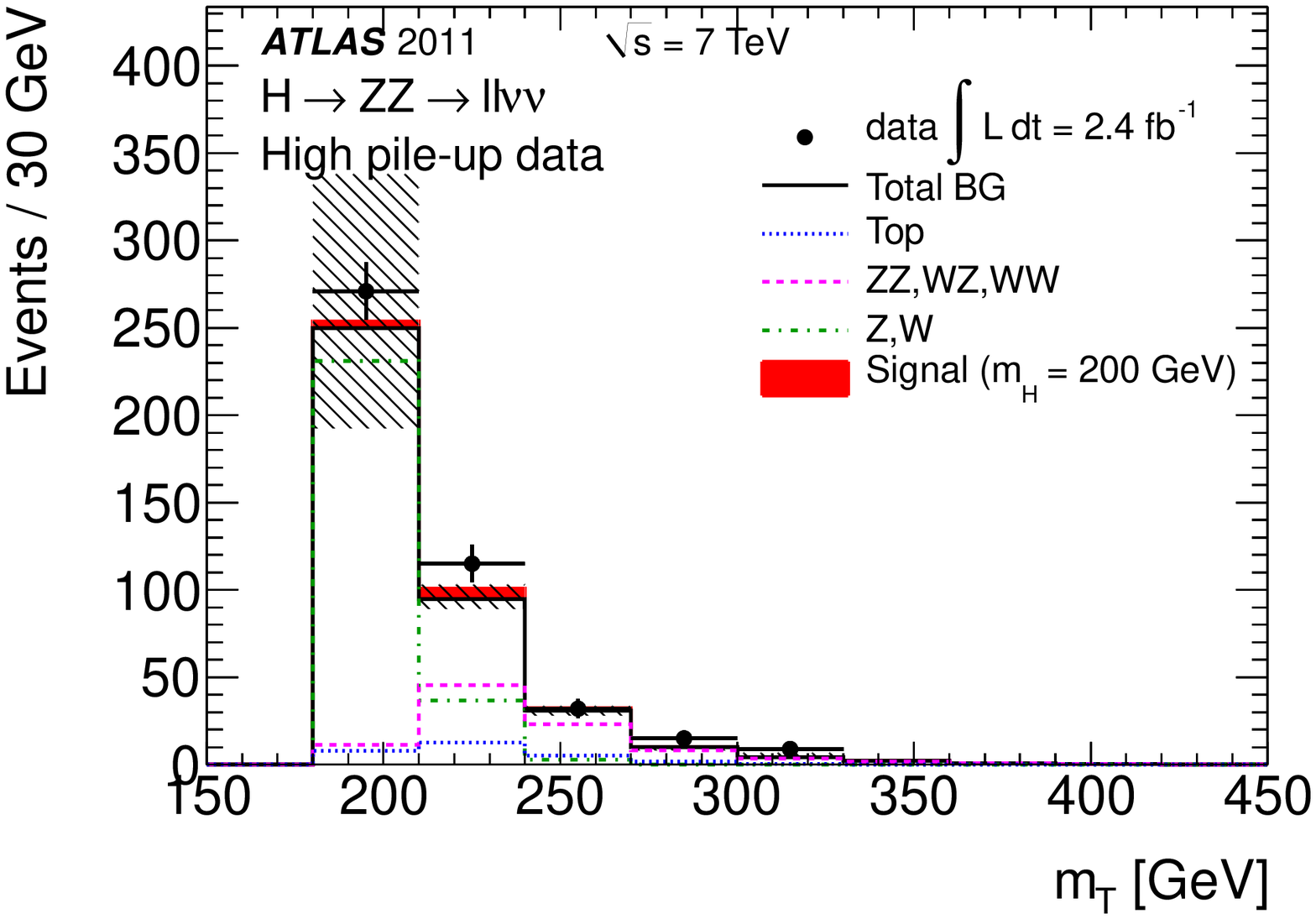}}
  \subfigure[\label{fig:htollvv_mass3}]{\includegraphics[width=0.49\textwidth]{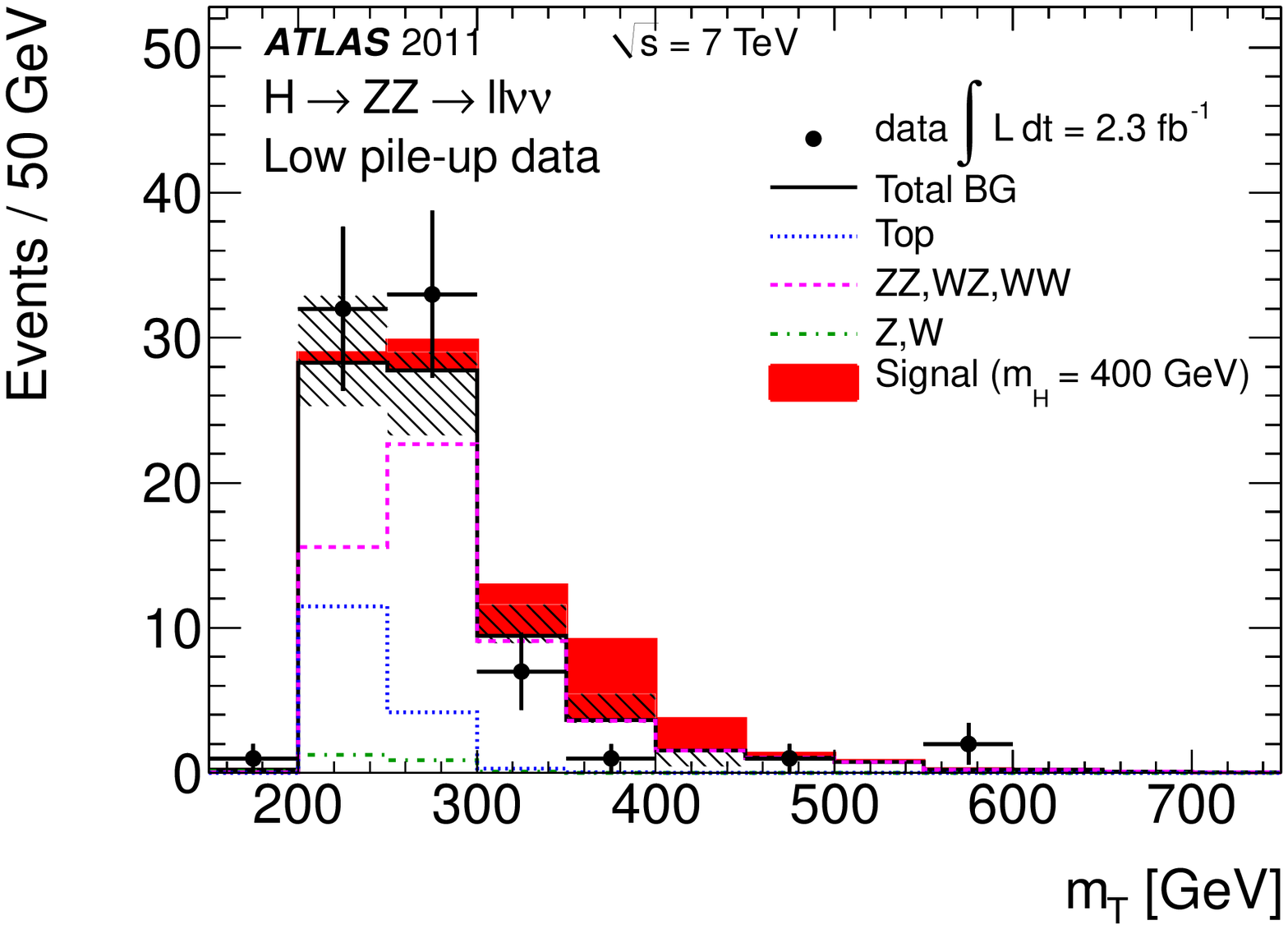}}
  \subfigure[\label{fig:htollvv_mass4}]{\includegraphics[width=0.49\textwidth]{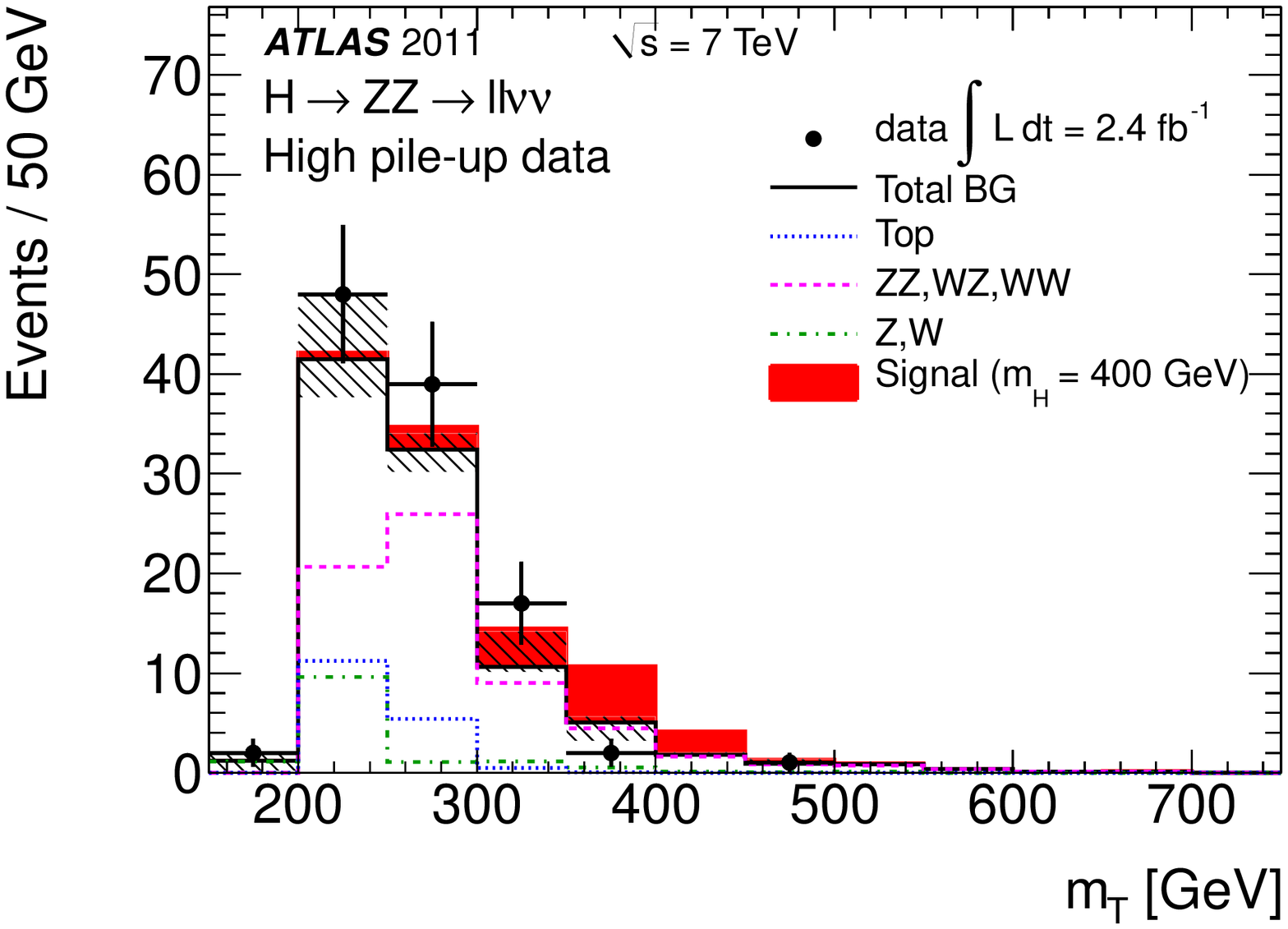}}
  \subfigure[\label{fig:htollvv_mass5}]{\includegraphics[width=0.49\textwidth]{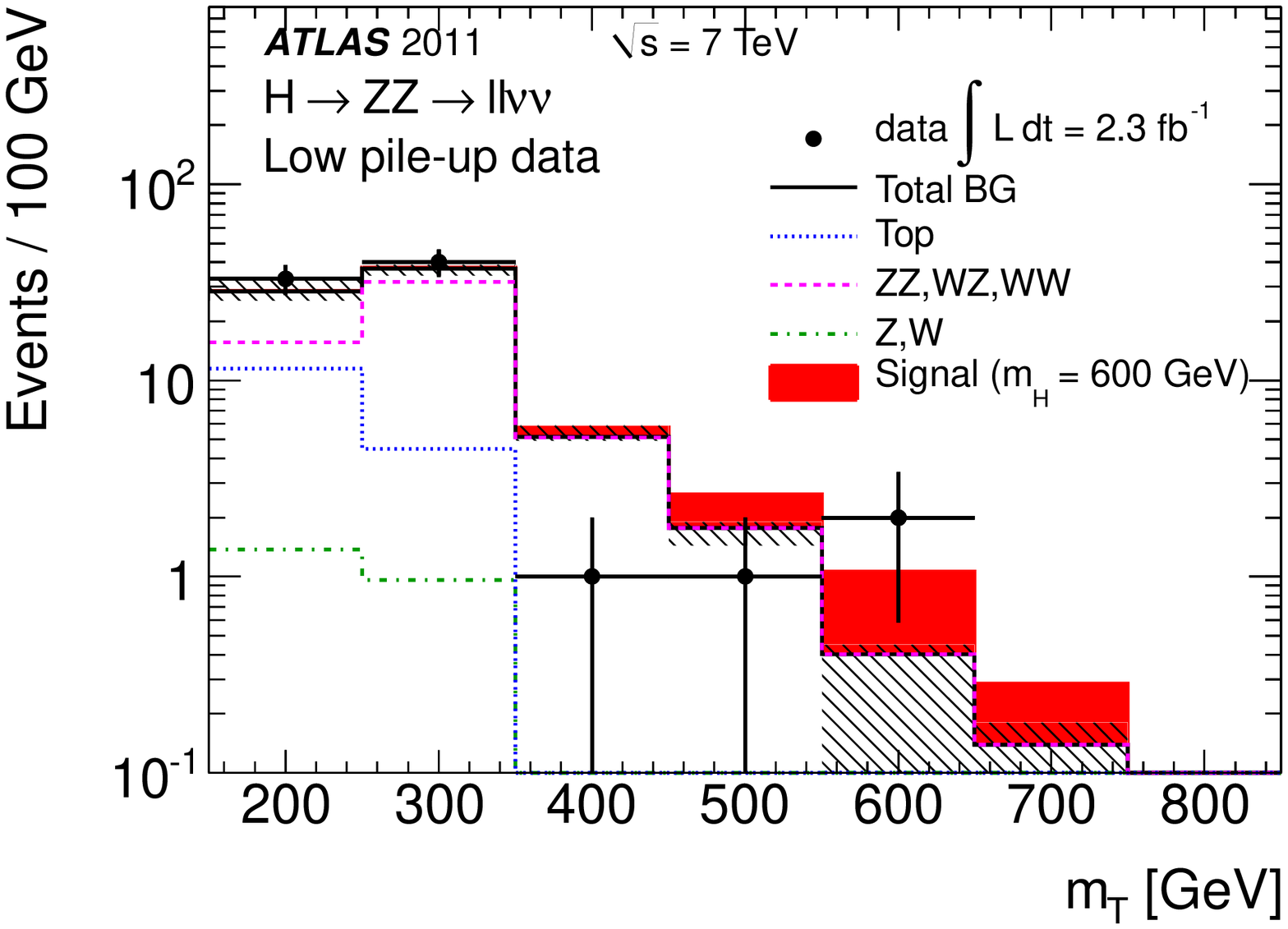}}
  \subfigure[\label{fig:htollvv_mass6}]{\includegraphics[width=0.49\textwidth]{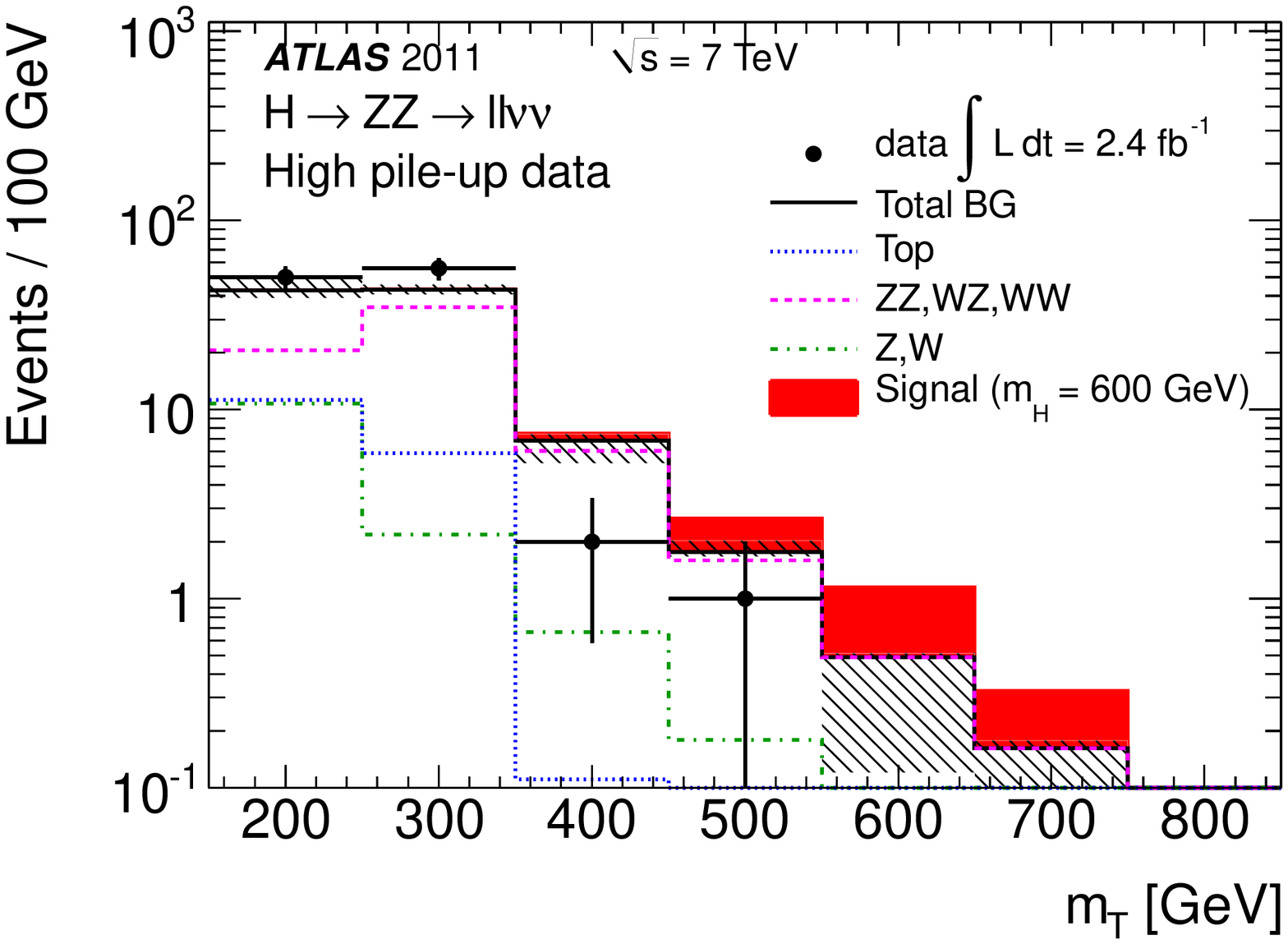}}
  \caption{\label{fig:htollvv_mass} The transverse mass distribution of \htollvv\ candidates. Figures \subref{fig:htollvv_mass1} and \subref{fig:htollvv_mass2} refer to the low \mH\ search region 
    and figures \subref{fig:htollvv_mass3} to \subref{fig:htollvv_mass6} to the high \mH~ search region. 
    The top, centre and bottom figures show the expected signal for a Higgs boson with a mass of 200~GeV, 400~GeV and 600~GeV, respectively. 
    Figures on the left correspond to the low pile-up data and figures on the right to the high pile-up data. The hashed area represents the combined systematic uncertainties on the total expected background.}           
\end{figure*}

Figure~\ref{fig:htollvv_mass} shows the $m_{\mathrm{T}}$ distributions in the low 
and high \mH\ search regions with the expected signal for a Higgs boson mass of 200~GeV, 
400~GeV and 600~GeV. The low pile-up and high pile-up data are shown separately.

The number and distribution of candidate \htollvv\ events observed in the data agree with the expected backgrounds. 
No indication of an excess is seen, with 
a smallest $p_0$-value of 0.05 at \mH=280~GeV, where $p_0$ represents the probablility that a background-only experiment would yield a result that is more signal-like than the observed result. 
Upper limits are set on the Higgs~boson production cross section relative to its predicted SM value as a function of \mH. 
The limits are extracted from a maximum likelihood fit to the $m_{\mathrm{T}}$ distribution following the $CL_s$ modified frequentist formalism~\cite{Read:2002hq} with 
the profile likelihood test statistic~\cite{Cowan:2010js}. All systematic uncertainties are taken into account.

\begin{figure*}
  \centering

 \includegraphics[width=0.7\textwidth]{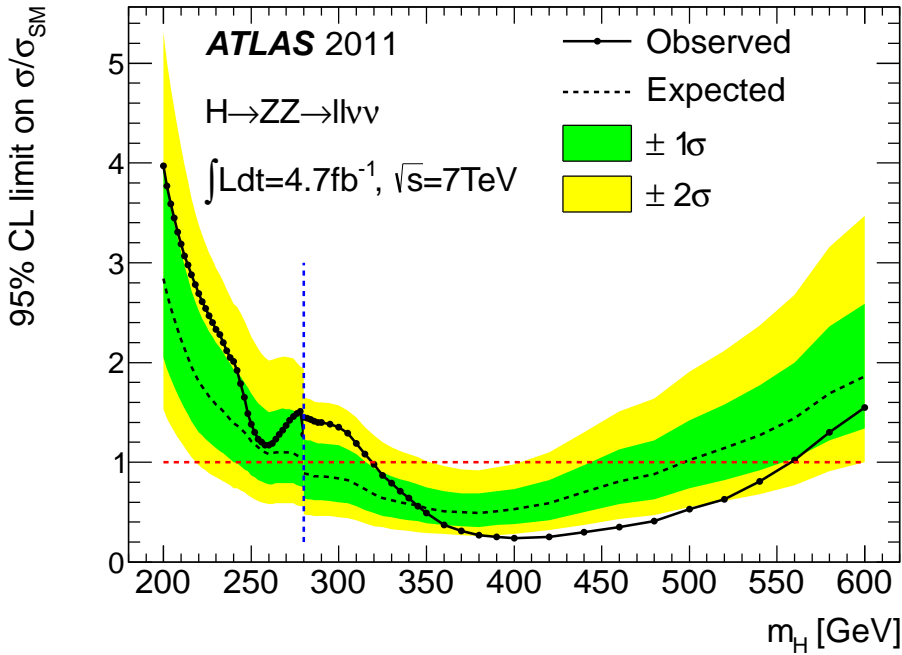}
  \caption{Observed and expected 95\% CL upper limits on the Higgs boson production cross~section divided by the SM prediction, 
    together with the green (inner) and yellow (outer) bands which indicate the $\pm 1\sigma$\, and $\pm 2 \sigma$\, fluctuations, respectively, around the median sensitivity. 
    The limits are based on \totallumi\ of data at $\sqrt{s}=$7~TeV. The line at 280 GeV indicates the transition from the low \mH\ to the high \mH\ search region.\label{fig:limits_llnunu_sig}}
\end{figure*}

Figure~\ref{fig:limits_llnunu_sig} shows the expected and observed limits at the 95\% CL. 
The green (inner) and yellow (outer) bands indicate the expected sensitivity with
$\pm 1\sigma$\, and $\pm 2 \sigma$\, 
fluctuations, respectively. 
The mass range between $280-497\:$~GeV is expected to be excluded while observation shows that a SM Higgs is excluded at the 95\% 
CL in the range of $319-558\:$~GeV.

\section{Summary}
Results of a search for a heavy SM Higgs~boson with a mass in the range $200 < \mH < 600$~GeV decaying to
 $ZZ\to\ell^+\ell^-\nu\bar{\nu}$ have been presented. These results are based on a data sample corresponding
to an integrated luminosity of \totallumi, recorded with the ATLAS detector at the LHC. No evidence for a signal is observed and 
cross section limits are placed over the mass range considered in this search, excluding the production of a SM Higgs~boson 
in the region $319 < \mH < 558$~GeV at the 95\% CL. 

%\clearpage

\section{Acknowledgments}

We thank CERN for the very successful operation of the LHC, as well as the
support staff from our institutions without whom ATLAS could not be
operated efficiently.

We acknowledge the support of ANPCyT, Argentina; YerPhI, Armenia; ARC,
Australia; BMWF, Austria; ANAS, Azerbaijan; SSTC, Belarus; CNPq and FAPESP,
Brazil; NSERC, NRC and CFI, Canada; CERN; CONICYT, Chile; CAS, MOST and NSFC,
China; COLCIENCIAS, Colombia; MSMT CR, MPO CR and VSC CR, Czech Republic;
DNRF, DNSRC and Lundbeck Foundation, Denmark; EPLANET and ERC, European Union;
IN2P3-CNRS, CEA-DSM/IRFU, France; GNAS, Georgia; BMBF, DFG, HGF, MPG and AvH
Foundation, Germany; GSRT, Greece; ISF, MINERVA, GIF, DIP and Benoziyo Center,
Israel; INFN, Italy; MEXT and JSPS, Japan; CNRST, Morocco; FOM and NWO,
Netherlands; RCN, Norway; MNiSW, Poland; GRICES and FCT, Portugal; MERYS
(MECTS), Romania; MES of Russia and ROSATOM, Russian Federation; JINR; MSTD,
Serbia; MSSR, Slovakia; ARRS and MVZT, Slovenia; DST/NRF, South Africa;
MICINN, Spain; SRC and Wallenberg Foundation, Sweden; SER, SNSF and Cantons of
Bern and Geneva, Switzerland; NSC, Taiwan; TAEK, Turkey; STFC, the Royal
Society and Leverhulme Trust, United Kingdom; DOE and NSF, United States of
America.

The crucial computing support from all WLCG partners is acknowledged
gratefully, in particular from CERN and the ATLAS Tier-1 facilities at
TRIUMF (Canada), NDGF (Denmark, Norway, Sweden), CC-IN2P3 (France),
KIT/GridKA (Germany), INFN-CNAF (Italy), NL-T1 (Netherlands), PIC (Spain),
ASGC (Taiwan), RAL (UK) and BNL (USA) and in the Tier-2 facilities
worldwide.

\bibliographystyle{elsarticle-num}
\bibliography{HIGG-2012-14}

\begin{thebibliography}{10}
\expandafter\ifx\csname url\endcsname\relax
  \def\url#1{\texttt{#1}}\fi
\expandafter\ifx\csname urlprefix\endcsname\relax\def\urlprefix{URL }\fi
\expandafter\ifx\csname href\endcsname\relax
  \def\href#1#2{#2} \def\path#1{#1}\fi

\bibitem{Englert:1964et}
F.~Englert, R.~Brout, {Broken Symmetry and the Mass of Gauge Vector Mesons},
  Phys. Rev. Lett. 13 (1964) 321--323.
\newblock \href{http://dx.doi.org/10.1103/PhysRevLett.13.321}
  {\path{doi:10.1103/PhysRevLett.13.321}}.

\bibitem{Higgs:1964pj}
P.~W. Higgs, {Broken Symmetries and the Masses of Gauge Bosons}, Phys. Rev.
  Lett. 13 (1964) 508--509.
\newblock \href{http://dx.doi.org/10.1103/PhysRevLett.13.508}
  {\path{doi:10.1103/PhysRevLett.13.508}}.

\bibitem{Guralnik:1964eu}
G.~S. Guralnik, C.~R. Hagen, T.~W.~B. Kibble, {Global Conservation Laws and
  Massless Particles}, Phys. Rev. Lett. 13 (1964) 585--587.
\newblock \href{http://dx.doi.org/10.1103/PhysRevLett.13.585}
  {\path{doi:10.1103/PhysRevLett.13.585}}.

\bibitem{Barate:2003sz}
{LEP Working Group for Higgs boson searches, ALEPH, DELPHI, L3 and OPAL
  Collaborations}, {Search for the standard model Higgs boson at LEP},
  Phys.~Lett. B~565 (2003) 61--75.
\newblock \href{http://arxiv.org/abs/hep-ex/0306033}
  {\path{arXiv:hep-ex/0306033}}.

\bibitem{Tevatron}
{CDF and D0 Collaborations}, {Combination of Tevatron searches for the standard
  model Higgs boson in the W+W- decay mode}, Phys. Rev. Lett. 104 (2010)
  061802.
\newblock \href{http://arxiv.org/abs/1001.4162} {\path{arXiv:1001.4162}}.

\bibitem{tevhiggs}
{The Tevatron New Physics and Higgs Working Group}, {Combined CDF and D0 Upper
  Limits on Standard Model Higgs Boson Production with up to 8.6~${\rm
  fb}^{-1}$ of Data} (2011).
\newblock \href{http://arxiv.org/abs/1107.5518} {\path{arXiv:1107.5518}}.

\bibitem{Collaboration:2012si}
{ATLAS Collaboration}, {Combined search for the Standard Model Higgs boson
  using up to 4.9~fb$^{-1}$ $p-p$ collisions at $\sqrt{s} = 7$~TeV with the
  ATLAS detector at the LHC}, Phys.~Lett. B~710 (2012) 49--66.
\newblock \href{http://arxiv.org/abs/1202.1408} {\path{arXiv:1202.1408}}.

\bibitem{Chatrchyan:2012tx}
{CMS Collaboration}, {Combined results of searches for the standard model Higgs
  boson in pp collisions at $\sqrt{s} = 7$~TeV}, Phys.~Lett. B~710 (2012)
  26--48.
\newblock \href{http://arxiv.org/abs/1202.1488} {\path{arXiv:1202.1488}}.

\bibitem{Aad:2011va}
{ATLAS Collaboration}, {Search for a Standard Model Higgs boson in the
  $H\rightarrow ZZ\rightarrow\ell^+\ell^-\nu\bar{\nu}$ decay channel with the
  ATLAS detector}, Phys.~Rev.~Lett. 107 (2011) 221802.
\newblock \href{http://arxiv.org/abs/1109.3357} {\path{arXiv:1109.3357}}.

\bibitem{Chatrchyan:2012vv}
{CMS Collaboration}, {Search for the standard model Higgs boson in the
  $H\rightarrow ZZ\rightarrow 2\ell 2\nu$ channel in pp collisions at $\sqrt{s}
  = 7$~TeV}, JHEP 03 (2012) 040.
\newblock \href{http://arxiv.org/abs/1202.3478} {\path{arXiv:1202.3478}}.

\bibitem{Aad:2011dr}
{ATLAS Collaboration}, {Luminosity Determination in $pp$ Collisions at
  $\sqrt{s}= 7\:{\rm TeV}$ Using the ATLAS Detector at the LHC}, Eur.~Phys.~J.
  C~71 (2011) 1630.
\newblock \href{http://arxiv.org/abs/1101.2185} {\path{arXiv:1101.2185}}.

\bibitem{lumi2011}
{ATLAS Collaboration}, \href{{http://cdsweb.cern.ch/record/1376384}}{Luminosity
  determination in $pp$ collisions at $\sqrt{s}=7\tev$ using the {ATLAS}
  detector in 2011}, {ATLAS-CONF-2011-116} (Aug 2011).
\newline\urlprefix\url{{http://cdsweb.cern.ch/record/1376384}}

\bibitem{Aad:2008zzm}
{ATLAS Collaboration}, {The ATLAS Experiment at the CERN Large Hadron
  Collider}, JINST 3 (2008) S08003.
\newblock \href{http://dx.doi.org/10.1088/1748-0221/3/08/S08003}
  {\path{doi:10.1088/1748-0221/3/08/S08003}}.

\bibitem{powheg4}
S.~Alioli, P.~Nason, C.~Oleari, E.~Re, {NLO Higgs boson production via gluon
  fusion matched with shower in POWHEG}, JHEP 04 (2009) 002.
\newblock \href{http://arxiv.org/abs/0812.0578} {\path{arXiv:0812.0578}}.

\bibitem{powheg5}
P.~Nason, C.~Oleari, {NLO Higgs boson production via vector-boson fusion
  matched with shower in POWHEG}, JHEP 02 (2010) 037.
\newblock \href{http://arxiv.org/abs/0911.5299} {\path{arXiv:0911.5299}}.

\bibitem{pythia}
T.~Sjostrand, S.~Mrenna, P.~Z. Skands, {PYTHIA 6.4 Physics and Manual}, JHEP 05
  (2006) 026.
\newblock \href{http://arxiv.org/abs/hep-ph/0603175}
  {\path{arXiv:hep-ph/0603175}}.

\bibitem{Golonka:2005pn}
P.~Golonka, Z.~Was, {PHOTOS Monte Carlo: A Precision tool for QED corrections
  in $Z$ and $W$ decays}, Eur.~Phys.~J. C~45 (2006) 97--107.
\newblock \href{http://arxiv.org/abs/hep-ph/0506026}
  {\path{arXiv:hep-ph/0506026}}.

\bibitem{Jadach:1993hs}
S.~Jadach, Z.~Was, R.~Decker, J.~H. Kuhn, {The tau decay library TAUOLA:
  Version 2.4}, Comput. Phys. Commun. 76 (1993) 361--380.
\newblock \href{http://dx.doi.org/10.1016/0010-4655(93)90061-G}
  {\path{doi:10.1016/0010-4655(93)90061-G}}.

\bibitem{Golonka:2003xt}
{P. Golonka \it{et al.}}, {The tauola-photos-F environment for the TAUOLA and
  PHOTOS packages, release II}, Comput. Phys. Commun. 174 (2006) 818--835.
\newblock \href{http://arxiv.org/abs/hep-ph/0312240}
  {\path{arXiv:hep-ph/0312240}}.

\bibitem{deFlorian:2011xf}
D.~de~Florian, G.~Ferrera, M.~Grazzini, D.~Tommasini, {Transverse-momentum
  resummation: Higgs boson production at the Tevatron and the LHC}, JHEP 11
  (2011) 064.
\newblock \href{http://arxiv.org/abs/1109.2109} {\path{arXiv:1109.2109}}.

\bibitem{ATLAS:2011aa}
{ATLAS Collaboration}, {Search for the Higgs boson in the $H\to WW^{(*)}\to
  \ell\nu\ell\nu$ decay channel in $pp$ collisions at $\sqrt{s} = 7\:{\rm TeV}$
  with the ATLAS detector}, Phys. Rev. Lett. 108 (2012) 111802.
\newblock \href{http://arxiv.org/abs/1112.2577} {\path{arXiv:1112.2577}}.

\bibitem{Aad:2011ec}
{ATLAS Collaboration}, {Search for a heavy Standard Model Higgs boson in the
  channel \htollqq\ using the ATLAS detector}, Phys.~Lett. B~707 (2012) 27--45.
\newblock \href{http://arxiv.org/abs/1108.5064} {\path{arXiv:1108.5064}}.

\bibitem{Collaboration:2012sm}
{ATLAS Collaboration}, {Search for the Standard Model Higgs boson in the decay
  channel $H\to ZZ^{(*)}\to 4\ell$ with 4.8~fb$^{-1}$ of $p-p$ collisions at
  $\sqrt{s}=7$~TeV with ATLAS}, Phys. Lett. B~710 (2012) 383--402.
\newblock \href{http://arxiv.org/abs/1202.1415} {\path{arXiv:1202.1415}}.

\bibitem{LHCHiggsCrossSectionWorkingGroup:2011ti}
{LHC Higgs Cross Section Working Group}, S.~Dittmaier, C.~Mariotti,
  G.~Passarino, R.~Tanaka~(Eds.), {Handbook of LHC Higgs cross sections: 1.
  Inclusive observables}, {CERN-2011-002} (2011).
\newblock \href{http://arxiv.org/abs/1101.0593} {\path{arXiv:1101.0593}}.

\bibitem{LHCHiggsCrossSectionWorkingGroup:2012vm}
{LHC Higgs Cross Section Working Group}, S.~Dittmaier, C.~Mariotti,
  G.~Passarino, R.~Tanaka~(Eds.), {Handbook of LHC Higgs Cross Sections: 2.
  Differential distributions}, {CERN-2012-002} (2012).
\newblock \href{http://arxiv.org/abs/1201.3084} {\path{arXiv:1201.3084}}.

\bibitem{Djouadi:1991tka}
A.~Djouadi, M.~Spira, P.~M. Zerwas, {Production of Higgs bosons in proton
  colliders: QCD corrections}, Phys. Lett. B~264 (1991) 440--446.
\newblock \href{http://dx.doi.org/10.1016/0370-2693(91)90375-Z}
  {\path{doi:10.1016/0370-2693(91)90375-Z}}.

\bibitem{Dawson:1990zj}
S.~Dawson, {Radiative corrections to Higgs boson production}, Nucl. Phys. B~359
  (1991) 283--300.
\newblock \href{http://dx.doi.org/10.1016/0550-3213(91)90061-2}
  {\path{doi:10.1016/0550-3213(91)90061-2}}.

\bibitem{Spira:1995rr}
M.~Spira, A.~Djouadi, D.~Graudenz, P.~M. Zerwas, {Higgs boson production at the
  LHC}, Nucl. Phys. B~453 (1995) 17--82.
\newblock \href{http://arxiv.org/abs/hep-ph/9504378}
  {\path{arXiv:hep-ph/9504378}}.

\bibitem{Harlander:2002wh}
R.~V. Harlander, W.~B. Kilgore, {Next-to-next-to-leading order Higgs production
  at hadron colliders}, Phys. Rev. Lett. 88 (2002) 201801.
\newblock \href{http://arxiv.org/abs/hep-ph/0201206}
  {\path{arXiv:hep-ph/0201206}}.

\bibitem{Anastasiou:2002yz}
C.~Anastasiou, K.~Melnikov, {Higgs boson production at hadron colliders in NNLO
  QCD}, Nucl. Phys. B~646 (2002) 220--256.
\newblock \href{http://arxiv.org/abs/hep-ph/0207004}
  {\path{arXiv:hep-ph/0207004}}.

\bibitem{Ravindran:2003um}
V.~Ravindran, J.~Smith, W.~L. van Neerven, {NNLO corrections to the total cross
  section for Higgs boson production in hadron hadron collisions}, Nucl. Phys.
  B~665 (2003) 325--366.
\newblock \href{http://arxiv.org/abs/hep-ph/0302135}
  {\path{arXiv:hep-ph/0302135}}.

\bibitem{Catani:2003zt}
S.~Catani, D.~de~Florian, M.~Grazzini, P.~Nason, {Soft-gluon resummation for
  Higgs boson production at hadron colliders}, JHEP 07 (2003) 028.
\newblock \href{http://arxiv.org/abs/hep-ph/0306211}
  {\path{arXiv:hep-ph/0306211}}.

\bibitem{Aglietti:2004nj}
U.~Aglietti, R.~Bonciani, G.~Degrassi, A.~Vicini, {Two-loop light fermion
  contribution to Higgs production and decays}, Phys.~Lett. B~595 (2004)
  432--441.
\newblock \href{http://arxiv.org/abs/hep-ph/0404071}
  {\path{arXiv:hep-ph/0404071}}.

\bibitem{Actis:2008ug}
S.~Actis, G.~Passarino, C.~Sturm, S.~Uccirati, {NLO Electroweak Corrections to
  Higgs Boson Production at Hadron Colliders}, Phys.~Lett. B~670 (2008) 12--17.
\newblock \href{http://arxiv.org/abs/0809.1301} {\path{arXiv:0809.1301}}.

\bibitem{Anastasiou:2008tj}
C.~Anastasiou, R.~Boughezal, F.~Petriello, {Mixed QCD-electroweak corrections
  to Higgs boson production in gluon fusion}, JHEP 04 (2009) 003.
\newblock \href{http://arxiv.org/abs/0811.3458} {\path{arXiv:0811.3458}}.

\bibitem{deFlorian:2009hc}
D.~de~Florian, M.~Grazzini, {Higgs production through gluon fusion: updated
  cross sections at the Tevatron and the LHC}, Phys.~Lett. B~674 (2009)
  291--294.
\newblock \href{http://arxiv.org/abs/0901.2427} {\path{arXiv:0901.2427}}.

\bibitem{Baglio:2010ae}
J.~Baglio, A.~Djouadi, {Higgs production at the lHC}, JHEP 03 (2011) 055.
\newblock \href{http://arxiv.org/abs/1012.0530} {\path{arXiv:1012.0530}}.

\bibitem{Ciccolini:2007jr}
M.~Ciccolini, A.~Denner, S.~Dittmaier, {Strong and electroweak corrections to
  the production of Higgs+2jets via weak interactions at the LHC}, Phys. Rev.
  Lett. 99 (2007) 161803.
\newblock \href{http://arxiv.org/abs/0707.0381} {\path{arXiv:0707.0381}}.

\bibitem{Ciccolini:2007ec}
M.~Ciccolini, A.~Denner, S.~Dittmaier, {Electroweak and QCD corrections to
  Higgs production via vector-boson fusion at the LHC}, Phys. Rev. D~77 (2008)
  013002.
\newblock \href{http://arxiv.org/abs/0710.4749} {\path{arXiv:0710.4749}}.

\bibitem{Arnold:2008rz}
{K. Arnold \it{et al.}}, {VBFNLO: A parton level Monte Carlo for processes with
  electroweak bosons}, Comput. Phys. Commun. 180 (2009) 1661--1670.
\newblock \href{http://arxiv.org/abs/0811.4559} {\path{arXiv:0811.4559}}.

\bibitem{Bolzoni:2010xr}
P.~Bolzoni, F.~Maltoni, S.-O. Moch, M.~Zaro, {Higgs production via vector-boson
  fusion at NNLO in QCD}, Phys. Rev. Lett. 105 (2010) 011801.
\newblock \href{http://arxiv.org/abs/1003.4451} {\path{arXiv:1003.4451}}.

\bibitem{Botje:2011sn}
{M. Botje \it{et al.}}, {The PDF4LHC working group interim recommendations}
  (2011).
\newblock \href{http://arxiv.org/abs/1101.0538} {\path{arXiv:1101.0538}}.

\bibitem{Lai:2010vv}
{H.-L.Lai, \it{et al.}}, {New parton distributions for collider physics}, Phys.
  Rev. D~82 (2010) 074024.
\newblock \href{http://arxiv.org/abs/1007.2241} {\path{arXiv:1007.2241}}.

\bibitem{Martin:2009iq}
A.~D. Martin, W.~J. Stirling, R.~S. Thorne, G.~Watt, {Parton distributions for
  the LHC}, Eur.~Phys.~J. C~63 (2009) 189--285.
\newblock \href{http://arxiv.org/abs/0901.0002} {\path{arXiv:0901.0002}}.

\bibitem{Ball:2011mu}
{R. D. Ball \it{et al.}}, {Impact of heavy quark masses on parton distributions
  and LHC phenomenology}, Nucl. Phys. B~849 (2011) 296--363.
\newblock \href{http://arxiv.org/abs/1101.1300} {\path{arXiv:1101.1300}}.

\bibitem{Djouadi:1997yw}
A.~Djouadi, J.~Kalinowski, M.~Spira, {HDECAY: A program for Higgs boson decays
  in the standard model and its supersymmetric extension}, Comput. Phys.
  Commun. 108 (1998) 56--74.
\newblock \href{http://arxiv.org/abs/hep-ph/9704448}
  {\path{arXiv:hep-ph/9704448}}.

\bibitem{Bredenstein:2006rh}
A.~Bredenstein, A.~Denner, S.~Dittmaier, M.~M. Weber, {Precise predictions for
  the Higgs-boson decay $H\rightarrow WW/ZZ \rightarrow 4$~leptons}, Phys. Rev.
  D~74 (2006) 013004.
\newblock \href{http://arxiv.org/abs/hep-ph/0604011}
  {\path{arXiv:hep-ph/0604011}}.

\bibitem{Bredenstein:2006ha}
A.~Bredenstein, A.~Denner, S.~Dittmaier, M.~M. Weber, {Radiative corrections to
  the semileptonic and hadronic Higgs-boson decays $H\rightarrow
  WW/ZZ\rightarrow 4$~fermions}, JHEP 02 (2007) 080.
\newblock \href{http://arxiv.org/abs/hep-ph/0611234}
  {\path{arXiv:hep-ph/0611234}}.

\bibitem{Seymour:1995qg}
M.~H. Seymour, {The Higgs boson line shape and perturbative unitarity}, Phys.
  Lett. B~354 (1995) 409--414.
\newblock \href{http://arxiv.org/abs/hep-ph/9505211}
  {\path{arXiv:hep-ph/9505211}}.

\bibitem{Passarino:2010qk}
G.~Passarino, C.~Sturm, S.~Uccirati, {Higgs pseudo-observables, second Riemann
  sheet and all that}, Nucl. Phys. B~834 (2010) 77--115.
\newblock \href{http://arxiv.org/abs/1001.3360} {\path{arXiv:1001.3360}}.

\bibitem{Anastasiou:2011pi}
C.~Anastasiou, S.~Buehler, F.~Herzog, A.~Lazopoulos, {Total cross-section for
  Higgs boson hadroproduction with anomalous Standard Model interactions}, JHEP
  12 (2011) 058.
\newblock \href{http://arxiv.org/abs/1107.0683} {\path{arXiv:1107.0683}}.

\bibitem{Mangano:2002ea}
{M. L. Mangano \it{et al.}}, {ALPGEN, a generator for hard multiparton
  processes in hadronic collisions}, JHEP 07 (2003) 001.
\newblock \href{http://arxiv.org/abs/hep-ph/0206293}
  {\path{arXiv:hep-ph/0206293}}.

\bibitem{Corcella:2000bw}
{G. Corcella \it{et al.}}, {HERWIG 6.5: an event generator for Hadron Emission
  Reactions With Interfering Gluons (including supersymmetric processes)}, JHEP
  01 (2001) 010.
\newblock \href{http://arxiv.org/abs/hep-ph/0011363}
  {\path{arXiv:hep-ph/0011363}}.

\bibitem{Frixione:2002ik}
S.~Frixione, B.~R. Webber, {Matching NLO QCD computations and parton shower
  simulations}, JHEP 06 (2002) 029.
\newblock \href{http://arxiv.org/abs/hep-ph/0204244}
  {\path{arXiv:hep-ph/0204244}}.

\bibitem{jimmy}
J.~M. Butterworth, J.~R. Forshaw, M.~H. Seymour, {Multiparton interactions in
  photoproduction at HERA}, Z. Phys. C~72 (1996) 637--646.
\newblock \href{http://arxiv.org/abs/hep-ph/9601371}
  {\path{arXiv:hep-ph/9601371}}.

\bibitem{Aad:2011qv}
{ATLAS Collaboration}, {Measurement of the production cross section for
  $Z/\gamma^*$ in association with jets in $pp$ collisions at $\sqrt{s} =
  7$~TeV with the ATLAS detector}, Phys. Rev. D~85 (2012) 032009.
\newblock \href{http://arxiv.org/abs/1111.2690} {\path{arXiv:1111.2690}}.

\bibitem{simuAtlas}
{ATLAS Collaboration}, {The ATLAS simulation Infrastructure}, Eur.~Phys.~J.
  C~70 (2010) 823--874.
\newblock \href{http://arxiv.org/abs/1005.4568} {\path{arXiv:1005.4568}}.

\bibitem{GEANT4}
{S. Agostinelli \it{et al.}}, {GEANT4: A simulation toolkit}, Nucl. Instrum.
  Meth. A~506 (2003) 250--303.
\newblock \href{http://dx.doi.org/10.1016/S0168-9002(03)01368-8}
  {\path{doi:10.1016/S0168-9002(03)01368-8}}.

\bibitem{Aad:2011mk}
{ATLAS Collaboration}, {Electron performance measurements with the ATLAS
  detector using the 2010 LHC proton-proton collision data}, Eur.~Phys.~J. C72
  (2012) 1909.
\newblock \href{http://arxiv.org/abs/1110.3174} {\path{arXiv:1110.3174}}.

\bibitem{Aad:2010yt}
{ATLAS Collaboration}, {Measurement of the $W\rightarrow \ell\nu$ and
  $Z/\gamma^* \rightarrow \ell\ell$ production cross sections in proton-proton
  collisions at $\sqrt{s}=7\;\tev$ with the ATLAS detector}, JHEP 12 (2010)
  060.
\newblock \href{http://arxiv.org/abs/1010.2130} {\path{arXiv:1010.2130}}.

\bibitem{AntiKt}
M.~Cacciari, G.~P. Salam, G.~Soyez, {The anti-$k_t$ jet clustering algorithm},
  JHEP 04 (2008) 063.
\newblock \href{http://arxiv.org/abs/0802.1189} {\path{arXiv:0802.1189}}.

\bibitem{jetcalibration}
{ATLAS Collaboration}, {Jet energy measurement with the ATLAS detector in
  proton-proton collisions at $\sqrt{s} = 7$~TeV}, {submitted to Eur.~Phys.~J}
  (2011).
\newblock \href{http://arxiv.org/abs/1112.6426} {\path{arXiv:1112.6426}}.

\bibitem{met}
{ATLAS Collaboration}, {Performance of missing transverse momentum
  reconstruction in proton-proton collisions at $\sqrt{s} = 7~\mbox{TeV}$ with
  ATLAS}, Eur.~Phys.~J. C~72 (2012) 1--35.
\newblock \href{http://arxiv.org/abs/1108.5602} {\path{arXiv:1108.5602}}.

\bibitem{btagnew}
{ATLAS Collaboration},
  \href{{https://cdsweb.cern.ch/record/1435197}}{{Measurement of the b-tag
  Effciency in a Sample of Jets Containing Muons with 5~fb$^{-1}$ of Data from
  the ATLAS Detector}}, {ATLAS-CONF-2012-043} (Mar 2012).
\newline\urlprefix\url{{https://cdsweb.cern.ch/record/1435197}}

\bibitem{mistagnew}
{ATLAS Collaboration},
  \href{{https://cdsweb.cern.ch/record/1435194}}{{Measurement of the Mistag
  Rate with 5~fb$^{-1}$ of Data Collected by the ATLAS Detector}},
  {ATLAS-CONF-2012-040} (Mar 2012).
\newline\urlprefix\url{{https://cdsweb.cern.ch/record/1435194}}

\bibitem{Campbell:2011bn}
J.~M. Campbell, R.~K. Ellis, C.~Williams, {Vector boson pair production at the
  LHC}, JHEP 07 (2011) 018.
\newblock \href{http://arxiv.org/abs/1105.0020} {\path{arXiv:1105.0020}}.

\bibitem{Binoth:2008pr}
{T. Binoth, N. Kauer and P. Mertsch}, {Gluon-induced QCD corrections to $pp \to
  ZZ \to \ell\bar{\ell}\ell'\bar{\ell'}$} (2008).
\newblock \href{http://arxiv.org/abs/0807.0024} {\path{arXiv:0807.0024}}.

\bibitem{auet2b}
{ATLAS Collaboration}, \href{{https://cdsweb.cern.ch/record/1400677}}{{Further
  ATLAS tunes of PYTHIA6 and Pythia 8}}, {ATL-PHYS-PUB-2011-014} (Nov 2011).
\newline\urlprefix\url{{https://cdsweb.cern.ch/record/1400677}}

\bibitem{Skands:2010ak}
P.~Z. Skands, {Tuning Monte Carlo Generators: The Perugia Tunes}, Phys. Rev.
  D82 (2010) 074018.
\newblock \href{http://arxiv.org/abs/1005.3457} {\path{arXiv:1005.3457}}.

\bibitem{Read:2002hq}
A.~L. Read, {Presentation of search results: The $CL_s$ technique}, J. Phys.
  G~28 (2002) 2693--2704.
\newblock \href{http://dx.doi.org/10.1088/0954-3899/28/10/313}
  {\path{doi:10.1088/0954-3899/28/10/313}}.

\bibitem{Cowan:2010js}
G.~Cowan, K.~Cranmer, E.~Gross, O.~Vitells, {Asymptotic formulae for
  likelihood-based tests of new physics}, Eur.~Phys.~J. C~71 (2011) 1554.
\newblock \href{http://arxiv.org/abs/1007.1727} {\path{arXiv:1007.1727}}.

\end{thebibliography}

\clearpage
\onecolumn
% ATLAS Collaboration author list for 21-FEB-2012
% Data extracted on 27-Apr-2012 for paperid 209
%\documentclass[11pt]{article}
%\usepackage{a4wide}\begin{document}
\begin{flushleft}
{\Large The ATLAS Collaboration}

\bigskip

G.~Aad$^{\rm 48}$,
B.~Abbott$^{\rm 111}$,
J.~Abdallah$^{\rm 11}$,
S.~Abdel~Khalek$^{\rm 115}$,
A.A.~Abdelalim$^{\rm 49}$,
O.~Abdinov$^{\rm 10}$,
B.~Abi$^{\rm 112}$,
M.~Abolins$^{\rm 88}$,
O.S.~AbouZeid$^{\rm 158}$,
H.~Abramowicz$^{\rm 153}$
% ...
\bigskip

% ...
$^{10}$ Institute of Physics, Azerbaijan Academy of Sciences, Baku, Azerbaijan\\
$^{11}$ Institut de F\'isica d'Altes Energies and Departament de F\'isica de la Universitat Aut\`onoma  de Barcelona and ICREA, Barcelona, Spain\\
% ...
$^{48}$ Fakult\"{a}t f\"{u}r Mathematik und Physik, Albert-Ludwigs-Universit\"{a}t, Freiburg i.Br., Germany\\
$^{49}$ Section de Physique, Universit\'e de Gen\`eve, Geneva, Switzerland\\
% ...
$^{88}$ Department of Physics and Astronomy, Michigan State University, East Lansing MI, United States of America\\
% ...
$^{111}$ Homer L. Dodge Department of Physics and Astronomy, University of Oklahoma, Norman OK, United States of America\\
$^{112}$ Department of Physics, Oklahoma State University, Stillwater OK, United States of America\\
% ...
$^{115}$ LAL, Universit\'e Paris-Sud and CNRS/IN2P3, Orsay, France\\
% ...
$^{153}$ Raymond and Beverly Sackler School of Physics and Astronomy, Tel Aviv University, Tel Aviv, Israel\\
% ...
$^{158}$ Department of Physics, University of Toronto, Toronto ON, Canada\\
% ...
\end{flushleft}

%\end{document} 

\end{document}
\bye